\documentstyle[12pt]{article}

\def\bs{\begin{subequations}}
\def\es{\end{subequations}}

\catcode`\@=11

\newtoks\@stequation
\def\subequations{\refstepcounter{equation}
  \edef\@savedequation{\the\c@equation}%
  \@stequation=\expandafter{\theequation}
  \edef\@savedtheequation{\the\@stequation}
  \edef\oldtheequation{\theequation}%
  \setcounter{equation}{0}%
  \def\theequation{\oldtheequation\alph{equation}}}

\def\endsubequations{\setcounter{equation}{\@savedequation}%
  \@stequation=\expandafter{\@savedtheequation}%
  \edef\theequation{\the\@stequation}\global\@ignoretrue}

\catcode`\@=12

\catcode`\@=11
\def\vereq#1#2{\lower3pt\vbox{\baselineskip1.5pt \lineskip1.5pt
\ialign{$\m@th#1\hfill##\hfil$\crcr#2\crcr\sim\crcr}}}
\catcode`\@=12

\makeatletter
        \renewcommand{\theequation}{\thesection.\arabic{equation}}%
        \@addtoreset{equation}{section}%
\makeatother

\begin{document}
\begin{titlepage}
\begin{center}
December 15, 1997     \hfill    UCB-PTH-97/63\\
                      \hfill    LBNL-41142\\

\vskip .25in

{\large \bf Asymptotic Search for Ground States \\ 
of $SU(2)$ Matrix Theory}

\vskip 0.3in

M.B. Halpern$^{a,b}$ and C. Schwartz$^a$

\vskip 0.15in

$^a${\em Department of Physics,
     University of California\\
     Berkeley, California 94720}\\
and\\
$^b${\em Theoretical Physics Group\\
     Ernest Orlando Lawrence Berkeley National Laboratory\\
     University of California,
     Berkeley, California 94720}
        
\end{center}

\vskip .3in

\vfill

\begin{abstract}
We introduce a complete set of gauge-invariant variables and a generalized 
Born-Oppenheimer formulation to search for normalizable zero-energy asymptotic 
solutions of the Schrodinger equation of $SU(2)$ matrix theory.  The asymptotic 
method gives only ground state candidates, which must be further tested for 
global stability.  Our results include a set of such ground state candidates, 
including one state which is a singlet under spin$(9)$.
\end{abstract}

\vfill

\end{titlepage}

\renewcommand{\thepage}{\arabic{page}}
\setcounter{page}{1}

\section{Introduction}

The $N=16$ supersymmetric gauge quantum mechanics 
\cite{Claud,Flume,Baake}, 
including its action formulation by dimensional reduction, was first 
studied in 1984-85. The model was noted again in 1988-89 as a regularization 
\cite{Wit}, with continuous spectrum \cite{Lusch}, of the $D=11$ supermembrane.  
In early 1996 the model was identified \cite{Witten} as the dynamics of 
interacting $D_0$-branes, which led to further study, including a truncated 
version of the model \cite{Dan} and the identification by $D_0$-scattering 
\cite{Kab,Doug} of the scale of $D_0$ physics with the scale of $D=11$ 
supergravity.

Interest in the $N=16$ model exploded in late 1996 when the large $n$ 
limit of the model, now christened Matrix theory, was proposed \cite{Bank} 
as a nonperturbative formulation of M-theory.  Among the many papers 
since then, we mention only the extension \cite{Suss} of the conjecture to
include finite $n$ and those papers with direct relevance to the ground 
state of the theory, in particular, the study of the Witten index of the 
theory \cite{Yi,Seth,Porr} and the ongoing study of the zero supercharge 
condition for SUSY ground states \cite{Hoppe,Hoppe2}.

Since the original work of Claudson and Halpern, however, the
ground state wave function of the theory has remained elusive. One 
obstruction to the investigation of such dynamical questions, pointed out 
in the original paper, is that matrix theory has no conserved fermion 
number, which  blocks the fermion sector analysis applicable to simpler 
supersymmetric quantum mechanical systems.  As a consequence, one expects 
that any particular matrix theory eigenstate is spread over a considerable 
portion of the fermionic Hilbert space.  The lore \cite{Witten,Bank} is that 
the theory should have a unique normalizable zero-energy ``threshold'' bound 
state, which is a singlet under spin$(9)$.

In this paper we develop an asymptotic method to search for zero-energy ground 
states of the $SU(2)$ matrix theory.  The method has two basic ingredients,

$\begin{array}{cl}
\bullet & \mbox{a complete set of gauge-invariant bosonic and fermionic 
variables} \\
\bullet & \mbox{a generalized Born-Oppenheimer formulation} 
\end{array}$

\noindent which allow us to extend some of the ideas of Ref.~\cite{Dan}.  
Moreover, there are strong parallels between our generalized Born-Oppenheimer 
formulation and the analysis of Ref.~\cite{Seth}. The method yields only 
candidate ground states, which are gauge-invariant asymptotic solutions, near 
the flat directions of the potential, of the zero-energy Schrodinger equation 
of the theory.  The ground state candidates must be further checked for global 
stability at non-asymptotic values of the gauge-invariant distance $R$.

Our results include a set of such candidate ground states, including 
exactly one state which is a singlet under spin$(9)$ and which, as it turns 
out, has bosonic angular momentum $l=2$.  The fermionic structure of the ground 
state candidates is relatively simple in the asymptotic domain, though one 
expects increasing complexity at higher order in $R^{-1}$.

\vspace{.4cm}
\noindent \underline{Matrix Theory}
\vspace{.3cm}  

We will follow the original notation \cite{Claud} for the theory, 
beginning with the 16 supercharges $Q_\alpha$,
\bs
\begin{equation} 
Q_\alpha = (\Gamma^m \Lambda_a)_\alpha \pi^m_a + igf_{abc} (\Sigma^{mn} 
\Lambda_a)_\alpha \phi^m_b \phi^n_c 
\label{1.1a}
\end{equation}
\begin{equation}
[\phi^m_a, \pi^n_b] = i\hbar \delta_{ab} \delta_{mn}, \; \; \; \; \; \; 
\{\Lambda_{a\alpha}, 
\Lambda_{b\beta}\} =  \delta_{ab} \delta_{\alpha\beta}
\label{1.1b}
\end{equation}
\begin{equation}
\{\Gamma^m, \Gamma^n\} = 2 \delta_{mn}, \; \; \; \; \; \; \Sigma^{mn} = 
-\frac{i}{4} [\Gamma^m, \Gamma^n] \\
\label{1.1c}
\end{equation}
\begin{equation}
a = 1 \ldots g, \; \; \; \; \; \; m = 1 \ldots 9, \; \; \; \; \; \; \alpha 
= 1 \ldots 16
\label{1.1d}
\end{equation}
\es
where $\phi^m_a$ are the real bosonic variables and $f_{abc}$ are the 
Cartesian structure constants of any compact Lie algebra with dimension $g$.  
The gamma matrices $(\Gamma^m)_{\alpha\beta}$ are real, symmetric and traceless 
and the fermions $\Lambda_{a\alpha}$ are real.  We will also need the generators 
$G_a$ of gauge transformations
\begin{equation}
G_a  =  f_{abc} (\phi^m_b \pi^m_c  -\frac{i\hbar}{2} \Lambda_{b\alpha} 
\Lambda_{c\alpha})
\label{1.2}
\end{equation}
and the generators $J^{mn}$ of spin$(9)$
\begin{equation}
J^{mn} = \pi_a^{[m} \phi_a^{n]} - \frac{\hbar}{2} \Lambda_{a\alpha} 
(\Sigma^{mn})_{\alpha\beta} 
\Lambda_{a\beta}
\label{1.3}
\end{equation}
where $[m \; n]$ means antisymmetrization of indices.

The supercharges satisfy
\begin{equation}
\{Q_\alpha, Q_\beta\}  = 2 \delta_{\alpha\beta} H + 2g  
(\Gamma^m)_{\alpha\beta}
\phi^m_a G_a 
\label{1.4}
\end{equation}
where H is the Hamiltonian
\bs
\begin{equation}
H = H_B + H_F
\label{1.5a}
\end{equation}
\begin{equation}
H_B = \frac{1}{2} \pi^m_a \pi^m_a  + V, \; \; \; \; \; \; V = \frac{g^2}{4} 
f_{abc} \phi^m_b \phi^n_c f_{ade} \phi^m_d \phi^n_e
\label{1.5b}
\end{equation}
\begin{equation}
H_F = - \frac{ig\hbar}{2} f_{abc} \Lambda_{a\alpha} (\Gamma^m)_{\alpha\beta} 
\phi^m_b 
\Lambda_{c\beta}
\label{1.5c}
\end{equation}   
\es                                                                                                    
and the gauge-invariant states $G_a \mid G.I. \, \rangle = 0$ form the 
physical subspace 
of the theory.

\section{Bosonic Preliminaries}

In this section we sharpen our tools on some bosonic subproblems, 
allowing the Lorentz vector index to run over $m = 1 \ldots d$ for 
generality, although $d=9$ for matrix theory.

\subsection{Gauge-Invariant Bosonic Variables}

For the gauge group $SU(2)$, with $f_{abc} = \epsilon_{abc}$, it is useful 
to define the real symmetric matrix $\Phi$
\begin{equation}
\Phi_{ab} \equiv \phi^m_a \phi^m_b, \; \; \; \; \; \; a,b = 1,2,3 	
\label{2.1}
\end{equation}
and the solutions to its eigenvalue problem
\bs
\begin{equation}
\Phi_{ab} \psi_b^i  =  \lambda_i^2 \psi_a^i, \; \; \; \; \; \;  i = 1,2,3
\label{2.2a}
\end{equation}
\begin{equation}
\psi_a^i \psi_a^j = \delta_{ij}, \; \; \; \; \; \; \psi_a^i \psi_b^i = 
\delta_{ab}	
\label{2.2b}
\end{equation}
\begin{equation}
\lambda_3 \geq \lambda_2 \geq \lambda_1 \geq 0.	
\label{2.2c}
\end{equation}
\es
The eigenvectors $\psi$ form a real orthogonal matrix and the eigenvalues 
$\lambda$ are a complete set of rotation- and gauge-invariant bosonic 
variables for this case.  

A complete set of $3(d-1)$ independent gauge-invariant bosonic 
variables includes the three eigenvalues $\lambda$ and the $3(d-2)$ 
gauge-invariant angular variables 
\begin{equation}
\eta^m_i \equiv \phi^m_a \psi_a^i /\lambda_i, \; \; \; \; \; \; \eta^m_i 
\eta^m_j = \delta_{ij}.	
\label{2.3}
\end{equation}
In the first part of this paper, we focus primarily on the gauge- and 
rotation-invariant $\lambda$'s, returning to the $\eta$'s in Section 7.  
For the gauge group $SU(3)$, there are more gauge-invariant variables, 
including $d_{abc} \phi_a^m \phi_b^n \phi_c^p$. 

On functions of $\lambda = (\lambda_1, \lambda_2, \lambda_3)$, the bosonic 
Hamiltonian takes the form
\bs
\begin{eqnarray}
H_B & = & -\frac{\hbar^2}{2} \Delta + V, \; \; \; \; \; \; \Delta = 
\partial_a^m \partial_a^m
\label{2.4a} \\
\Delta f(\lambda) & = & \rho^{-1} \frac{\partial}{\partial \lambda_i} 
(\rho \frac{\partial}{\partial \lambda_i}
f(\lambda))	
\label{2.4b} \\
V & = & \frac{g^2}{2} (\lambda_1^2 \lambda_2^2 + \lambda_2^2 \lambda_3^2 + 
\lambda_3^2 
\lambda_1^2)	
\label{2.4c} \\
\rho(\lambda) & \equiv & (\lambda_1 \lambda_2 \lambda_3)^{d-3} 
(\lambda_3^2 - \lambda_1^2) 
(\lambda_3^2 - \lambda_2^2) (\lambda_2^2 - \lambda_1^2) \; \geq 0
\label{2.4d}
\end{eqnarray}
\es
where $\partial_a^m = \partial/\partial \phi_a^m$ and $H_B$ is 
hermitian in the inner product
\begin{equation}
\int d^3\lambda \rho(\lambda) f^\ast(\lambda) g(\lambda), \; \; \; \; \; 
\;     
d^3\lambda \equiv d\lambda_1 d\lambda_2 d\lambda_3. 
\label{2.5}
\end{equation}
More generally, the full bosonic measure is
\bs
\begin{eqnarray}
(d\phi) & = & d^3\lambda \rho(\lambda) (d\Omega)	
\label{2.6a} \\
\int (d\Omega) & = & 1	
\label{2.6b} \\
\int (d\Omega) f(\phi) & = & 0 \; \; \; \; \; \; {\rm when} \; f(\phi) 
= -f(-\phi)	
\label{2.6c}
\end{eqnarray}
\es
where $\Omega$ are $3(d-1)$ ``angles'' (which include the $3(d-2)$ 
gauge-invariant angles $\eta$ in (2.3), plus three gauge degrees of freedom).  
Through Section 6 of this paper, the relations (2.6b) and (2.6c) are all we 
shall need to know about $\Omega$.

\subsection{Zero-Energy Hamilton-Jacobi Equation} 

It was emphasized by Claudson and Halpern that a SUSY ground state 
must satisfy the zero-energy Hamilton-Jacobi equation
\bs
\begin{equation}
\psi \mathop{\sim}\limits_{\hbar \rightarrow 0} \exp[\pm \frac{S_0}{\hbar}]
\label{2.7a}
\end{equation}
\begin{equation}
\frac{1}{2} \mid \bigtriangledown S_0 \mid^2 = V	
\label{2.7b}
\end{equation}
\es
in the extreme semiclassical limit, and this equation takes the 
$d$-independent form
\begin{equation}
(\partial/\partial\lambda_i S_0(\lambda)) (\partial/\partial\lambda_i 
S_0(\lambda)) =  
g^2(\lambda_1^2 \lambda_2^2 + \lambda_2^2 \lambda_3^2 + \lambda_3^2 
\lambda_1^2)
\label{2.8}
\end{equation}
when we restrict ourselves to gauge- and rotation-invariant wave 
functions.  These authors also gave an exact solution of eq.~(2.7b) or 
(2.8),
\bs
\begin{eqnarray}
S_0(\lambda) & = & \sqrt{W} 
\label{2.9a} \\
W & = & \frac{1}{6} g^2 \epsilon_{abc} \epsilon_{def} \phi_a^m \phi_b^n 
\phi_c^p \phi_d^m \phi_e^n \phi_f^p 	
\label{2.9b} \\
& = & g^2 det \Phi 
\label{2.9c} \\
& = & g^2 \lambda_1^2 \lambda_2^2 \lambda_3^2.
\label{2.9d}
\end{eqnarray}	
\es
Here, we make a brief systematic study of the solutions of the zero-energy 
Hamilton-Jacobi equation (2.8).

For this investigation, it is convenient to introduce spherical 
coordinates
\bs
\begin{equation}
\lambda_1 = r \sin \theta \cos \phi, \; \; \; \; \; \; \lambda_2 = r \sin 
\theta \sin \phi, 
\; \; \; \; \; \; \lambda_3 = r \cos \theta	
\label{2.10a}
\end{equation}
\begin{equation}
r^2 = \lambda_1^2 + \lambda_2^2  + \lambda_3^2 	
\label{2.10b}
\end{equation}
\es
and to write the solution in the form
\begin{equation}
S_0 = g r^3 F(\theta,\phi).
\label{2.11}
\end{equation}
Then the Hamilton-Jacobi equation (2.8) reduces to
\begin{equation}
\sin^2\theta (9 F^2 + F_\theta^2) + F_\phi^2 =  \sin^4\theta (\cos^2\theta 
+ \sin^2\theta \cos^2\phi \sin^2\phi)	
\label{2.12}
\end{equation}
where the subscripts denote partial derivatives.  The right side of (2.12) 
is proportional to $V$, so the flat directions of $V$ correspond to 
$\theta = 0$ in these variables.

For small $\theta$, (2.12) admits two solutions which are non-singular as 
$\theta \rightarrow 0$,
\begin{equation}
S_0 = \frac{g r^3 \theta^2}{2} 
\left\{ \begin{array}{c}
\sin2(\phi - \phi_0) \\
1 \end{array} \right\}
+ O(\theta^3).	
\label{2.13}
\end{equation}
The first solution, which we call the Claudson-Halpern (CH) branch, 
contains the CH solution $S_0(\lambda) = \sqrt{W}$ when $\phi_0 = 0$, and 
the second solution is the solution $S_0(\lambda) \cong \frac{V}{gr}$ 
studied later by Itoyama \cite{Ito,Razz}.  The  $\phi_0 \neq 0$ solutions
of the CH branch are new.

Because the exponential decrease of $\exp[-\frac{S_0}{\hbar}]$ at large 
$r$ is lost near $\theta=0$, none of these solutions is normalizable\footnote{We 
note in passing that the full bosonic Hamiltonian $H_B$ has exact gauge-invariant
zero-energy solutions 
\[ \psi(W) = W^\gamma K_{2\mid \gamma \mid} (\frac{\sqrt{W}}{\hbar}) 
\mathop{\sim}\limits_{\hbar \rightarrow 0} \exp[- \frac{\sqrt{W}}{\hbar}], 
\; \; \; \; \; \; \gamma = \frac{(4-d)}{4} 
\]
and $K \rightarrow I$ where $K$ and $I$ are cylinder functions of 
imaginary argument.  These solutions are quantum extensions of the 
Claudson-Halpern solution, but neither are normalizable.}, whether we choose 
the naive measure $d^3\lambda$ or the quantum measure $d^3\lambda \rho(\lambda)$.  
More precisely, we find non-normalizability in the flat directions
\bs
\begin{equation}
\int d^3\lambda \exp[- \frac{2 S_0}{\hbar}] \; \; \; \; \; \propto 
\; \; \; \; \; \int_0 \frac{d\theta}{\theta} 	
\label{2.14a}
\end{equation}
\begin{equation}
\int d^3\lambda \rho(\lambda) \exp[- \frac{2 S_0}{\hbar}] \; \; \; \; \; 
\; \propto \; \; \; \; \; \; \int_0 \frac{d\theta}{\theta^3}	
\label{2.14b}
\end{equation}
\es
for the Itoyama solution and the CH branch with $0 < \phi_0 < \frac{\pi}{4}$.
(For small $\theta$ the range of the angle $\phi$ is $\frac{\pi}{4} \leq 
\phi \leq \frac{\pi}{2}$.)  The CH solution itself, with $\phi_0 = 0$, has an extra 
multiplicative divergence in the $\phi$ integration.

It is clear that the Hamilton-Jacobi equation by itself is unable to choose 
among solutions or to answer the question of normalizability:  any of the solutions might, in 
principle, be made normalizable by quantum corrections including the 
fermions, and
\begin{equation}
\psi \sim \exp[- \frac{1}{\hbar} (S_0 + \hbar \alpha \ln r)]	
\label{2.15}
\end{equation}
is normalizable in $d^3\lambda \rho(\lambda)$ when $\alpha > \frac{3}{2}$ 
for the Itoyama solution and for the CH branch with $0 < \phi_0 < \frac{\pi}{4}$.
For the CH solution itself, normalizability requires that $\alpha > 3$.  In 
what follows, our task is to study such quantum corrections in detail.

It is also important to note that our study of the zero-energy 
Hamilton-Jacobi equation is incomplete because the full equation (2.7b) 
allows other (gauge- but not rotation-invariant) solutions, which include 
the $\eta$ variables in (2.3)  as well (see Section 7).

\subsection{Born-Oppenheimer Approximation}

Our approach in this paper follows the line of the Born-Oppenheimer 
approximation \cite{Born}, which we illustrate first on the gauge- and 
rotation-invariant sector of the bosonic Hamiltonian
\bs
\begin{equation}
H_B \psi(\lambda) = E \psi(\lambda)	
\label{2.16a}
\end{equation}
\begin{equation}
V = \frac{g^2}{2} [R^2 (\lambda_1^2 + \lambda_2^2) + \lambda_1^2 
\lambda_2^2]
\label{2.16b}
\end{equation}
\es
where we have set $R = \lambda_3$.  Our goal is to study the asymptotic 
behavior at large $R$ near the classical flat directions, $\lambda_1 = 
\lambda_2 = 0$, of $V$.  In the language of the Born-Oppenheimer approximation, 
we integrate out the ``fast'' variables $\lambda_1, \lambda_2$ to obtain an 
effective Hamiltonian for the asymptotic behavior in the ``slow'' variable 
$R = \lambda_3$.

Toward this end we first decompose the Hamiltonian and the measure as
\bs
\begin{equation}
H_B  = H_0 + H_1	
\label{2.17a}
\end{equation}
\begin{eqnarray}
H_0 & = & -\frac{\hbar^2}{2} \{ \frac{\partial^2}{\partial\lambda_1^2} + 
[\frac{(d-3)}{\lambda_1} + \frac{2\lambda_1}{(\lambda_1^2 - \lambda_2^2)}] 
\frac{\partial}{\partial\lambda_1} + \frac{\partial^2}{\partial\lambda_2^2} 
\nonumber \\
& & + [\frac{(d-3)}{\lambda_2} + \frac{2\lambda_2}{(\lambda_2^2 - \lambda_1^2)}] 
\frac{\partial}{\partial\lambda_2} \} + \frac{g^2}{2} R^2 (\lambda_1^2 + 
\lambda_2^2)
\label{2.17b}
\end{eqnarray}
\begin{eqnarray}
H_1 & = & -\frac{\hbar^2}{2} \{\frac{\partial^2}{\partial R^2} + [\frac{(d+1)}
{R} + \frac{2(\lambda_1^2 + \lambda_2^2)}{R^3} + \ldots] \frac{\partial}
{\partial R} \nonumber \\
& & - \frac{2}{R^2} (\lambda_1 \frac{\partial}{\partial\lambda_1} + 
\lambda_2 \frac{\partial}{\partial\lambda_2} + \ldots)\} 
+ \frac{g^2}{2} \lambda_1^2 \lambda_2^2
\label{2.17c} \\
\rho & = & R^{d+1} \sigma	
\label{2.17d} \\
\sigma & \equiv & (\lambda_1 \lambda_2)^{d-3} (\lambda_2^2 - \lambda_1^2) 
(1 - \frac{\lambda_1^2}{R^2}) (1 - \frac{\lambda_2^2}{R^2}) 
\label{2.17e}
\end{eqnarray}
\es
where we will see that $H_0$ is the dominant part of $H_B$ at large $R$ and
the dots in $H_1$ indicate terms with higher inverse powers of $R$.

The first term $H_0$ in (2.17b) describes a rotation- and gauge-invariant 
two-dimensional oscillator whose frequency is linear in $R$.  The nodeless
eigenstate
\bs
\begin{equation}
u_R (\lambda_1, \lambda_2) = C(R) R^{\frac{(d-1)}{2}} 
\exp[-(\frac{gR}{2\hbar}) (\lambda_1^2 + 
\lambda_2^2)]
\label{2.18a}
\end{equation}
\begin{equation}
E_0(R) = \hbar g R (d-1)	
\label{2.18b}
\end{equation}
\begin{equation}
\int d^2\lambda \: \sigma \mid u_R \mid^2 = 1	
\label{2.18c}
\end{equation}
\es
is almost certainly the unique ground state of $H_0$ (see Appendix G), 
where $E_0(R)$ is the energy and $d^2\lambda = d\lambda_1 d\lambda_2$.  
The power of $R$ in $u_R$ guarantees that $C(R)$ approaches a constant 
at large $R$,
\begin{equation}
\mid C(R) \mid^2 = \frac{(d-3)!}{2^{d-1}} (1 + O(R^{-3}))	
\label{2.19}
\end{equation}
because
\begin{equation}
\lambda_1, \lambda_2 = O((\frac{\hbar}{g})^\frac{1}{2} R^{-\frac{1}{2}})	
\label{2.20}
\end{equation}
when averaged over $\mid u_R \mid^2$.  The orders of magnitude in (2.20)
define the quantum neighborhood of the classical flat directions of the 
potential.

In this paper we compute only through $O(R^{-2})$, and, for this purpose, 
$C(R)$ may be treated as a constant.  Similarly, the measure $\sigma$ in 
(2.17e) can be replaced by its asymptotic form
\begin{equation}
\sigma \rightarrow \sigma_\infty = (\lambda_1 \lambda_2)^{d-3} 
(\lambda_2^2 - \lambda_1^2)	
\label{2.21}
\end{equation}	   
in all computations through $O(R^{-2})$. 

More generally, all the eigenfunctions of $H_0$ can be written as the 
normalizing power of $R$ in (2.18a) times functions of the scaled variables
\begin{equation}
z_1 = \lambda_1 (\frac{g}{\hbar}R)^\frac{1}{2}, \; \; \; \; \; \; 
z_2 = \lambda_2 (\frac{g}{\hbar}R)^\frac{1}{2}	
\label{2.22}
\end{equation}  
and the corresponding eigenvalues are all proportional to $R$.  At finite
values of $z_{1,2}$, it follows 
that, throughout the Hilbert space of $H_0$, we may estimate the order of 
magnitude of $\lambda_1$ or $\lambda_2$ at large $R$ as $O(R^{-\frac{1}{2}})$
(as recorded in (2.20)), 
and the derivatives with respect to $\lambda_1$ or $\lambda_2$ as 
$O(R^{\frac{1}{2}})$.  Using these orders of magnitude, one sees that 
$H_0$ is the dominant part of $H_B$ (contains all terms of $O(R)$) in the
gauge- and rotation-invariant sector and $H_1 = O(R^{-2})$.

Another way to view the large $R$ expansion of this paper, although
we have chosen not to write things out in this way, is to use as  
independent variables ($z_1, z_2, \lambda_3 = R$), and then formally 
expand in powers of $R^{-1}$.

The conventional Born-Oppenheimer approximation is essentially 
first-order perturbation theory in $H_1$ around $u_R$.  In variational 
language, we study a separable trial wave function of the form
\begin{equation}
\psi(\lambda) = u_R(\lambda_1, \lambda_2) \psi(R)	
\label{2.23}
\end{equation}
where $\psi(R)$ may be called the reduced wave function or state vector.  
Averaging over the fast variables, we obtain an effective Hamiltonian for 
the slow variable $R$,
\bs
\begin{eqnarray}
H_{eff} (R) \psi(R) & = & E \psi(R)	
\label{2.24a} \\
H_{eff} (R) & = & \int d^2\lambda \sigma u_R^\ast H_B u_R	
\label{2.24b}
\end{eqnarray}
\vspace{-12pt}
\begin{equation}
\int dR R^{d+1} \mid \psi(R) \mid^2 \; < \infty	
\label{2.24c}
\end{equation}
\es
where the normalization condition on the reduced state vector is given in 
(2.24c).  The effective Hamiltonian (2.24b) can be evaluated exactly but we 
confine ourselves in this paper to the leading terms (through $O(R^{-2})$)
at large $R$.  

Using the integrals given in Appendix F, we obtain the asymptotic form
of the effective Hamiltonian
\bs
\begin{equation}
H_{eff} (R) = g\hbar(d-1)R - \frac{\hbar^2}{2} \{\frac{d^2}{dR^2} + 
\frac{(d+1)}{R} \frac{d}{dR} + \frac{B}{R^2} + \ldots\}
\label{2.25a}
\end{equation}
\begin{equation}
B = -\frac{(d-1)(d-9)}{4}
\label{2.25b}
\end{equation}
\es
whose linear potential is nothing but $E_0(R)$ in (2.18b).  The coefficient of 
the first derivative term in (2.25a) could have been fixed in advance by 
hermiticity of $H_{eff}$ in the reduced measure $R^{d+1}$ of eq.~(2.24c) 
and in fact the operator
\begin{equation}
\Delta _R = \frac{d^2}{dR^2} + \frac{(d+1)}{R} \frac{d}{dR}	
\label{2.26}
\end{equation}
is the natural Laplacian on this measure.

For this bosonic case, the positive potential growing linearly with $R$ 
gives $E > 0$ normalizable bound states which show exponential decrease
\begin{equation}
\psi(R) \sim \exp[-\frac{2}{3} (\frac{2g(d-1)}{\hbar})^{\frac{1}{2}} 
R^{\frac{3}{2}}]
\label{2.27}
\end{equation}
at large $R$.  For the full matrix theory, we expect from Ref.~\cite{Dan} that 
the fermionic contributions will exactly cancel\footnote{Following Ref.~\cite{Dan}, 
we expect that sectors with uncancelled linear $R$ terms are associated to 
excited states.} the bosonic contribution $E_0(R) = 8g\hbar R$, leaving an 
effective Hamiltonian of the form
\begin{equation}
H_{eff} (R) = - \frac{\hbar^2}{2} (\frac{d^2}{dR^2} + \frac{(d+1)}{R} 
\frac{d}{dR} + \frac{B^\prime}{R^2} + \ldots)	
\label{2.28}
\end{equation}
and such a Hamiltonian can have an asymptotic power-law behaved 
zero-energy normalizable bound state, provided that
\begin{equation}
B^\prime < \frac{(d^2-4)}{4}.
\label{2.29}
\end{equation}

The problem here is that the Born-Oppenheimer approximation cannot be 
trusted to give the true value of the constant $B^\prime$, even 
approximately, because (unlike molecular physics) matrix theory has 
no natural small parameters to control the approximation.  In what 
follows, we develop an improved formalism which allows us to compute 
the necessary coefficient $B^\prime$ exactly in matrix theory.

\section{Generalized Perturbation Theory} 

In order to study the asymptotic behavior of the wavefunction, we need 
a procedure which combines the idea of the Born-Oppenheimer approximation 
with the techniques of perturbation theory.  Here is such a general 
formalism for studying the equation
\begin{equation}
{\cal L} \mid \Psi \, \rangle = 0  
\label{3.1}
\end{equation}
in which the linear operator is ${\cal L} = H-E$ and $\mid \Psi \, \rangle$ 
is a vector in the Hilbert space of $H$.

We start by choosing a normalized state $\mid  {\bf \cdot} \, \rangle$ in the Hilbert 
space and its associated projection operators
\begin{equation}
P = P^2 = \; \mid  {\bf \cdot} \, \rangle \langle \, {\bf \cdot} \mid,  \; \; \; \; \; \;  Q = Q^2 = 1 
- P 
\label{3.2}
\end{equation}
and the action of these projection operators on the state vector will be 
written as 
\begin{equation}
\mid \Psi_P \, \rangle = P \mid \Psi \, \rangle, \; \; \; \; \; \; \mid \Psi_Q 
\, \rangle = Q \mid \Psi \, \rangle.  
\label{3.3}
\end{equation}

The original Schrodinger equation (3.1) is then broken down into two 
coupled equations. The first equation is
\begin{equation}
P {\cal L} P \mid \Psi_P \, \rangle + P {\cal L} Q \mid \Psi_Q \, \rangle = 0 
\label{3.4}
\end{equation}
or, equivalently,
\begin{equation}
\langle \, {\bf \cdot} \mid {\cal L} \mid  {\bf \cdot} \, \rangle \langle \, {\bf \cdot} 
\mid \Psi_P \, \rangle + \langle \, {\bf \cdot} \mid {\cal L} Q \mid \Psi_Q \, 
\rangle = 0 
\label{3.5}
\end{equation}
and the second equation is
\begin{equation}
Q {\cal L} P \mid \Psi_P \, \rangle + Q {\cal L}Q \mid \Psi_Q\, \rangle = 0   
\label{3.6}
\end{equation}
which can be formally solved as,
\begin{equation}
\mid \Psi_Q \, \rangle = - (Q {\cal L} Q)^{-1} Q {\cal L} P \mid 
\Psi_P \, \rangle.  
\label{3.7}
\end{equation}
If we substitute (3.7) into (3.4), we get the ``reduced'' Schrodinger 
equation:
\begin{equation}
[P {\cal L} P - P {\cal L} Q (Q {\cal L} Q)^{-1} Q {\cal L} P] \mid \Psi_P 
\, \rangle = 0 
\label{3.8}
\end{equation}
or, equivalently,
\begin{equation}
[\langle \, {\bf \cdot} \mid {\cal L} \mid  {\bf \cdot} \, \rangle - \langle \, {\bf \cdot} 
\mid {\cal L} Q (Q {\cal L} Q)^{-1} Q {\cal L} \mid  {\bf \cdot} \, \rangle]
\langle \,{\bf \cdot} \mid \Psi \, \rangle = 0.
\label{3.9}
\end{equation}
One can also write a variational principle for the exact solution of (3.6),
\begin{equation}
J[\chi] = \langle \,\chi\mid Q{\cal L} Q \mid\chi\, \rangle + 2 \langle \, \chi
\mid Q{\cal L} P \mid \Psi_P \, \rangle
\label{3.10}
\end{equation}
where $J$ is stationary under variations of $\mid \chi \, \rangle$ about
$\mid \Psi_Q \, \rangle$.

This formulation is exact and can be adapted to a number of different 
applications.  For the familiar problem of non-degenerate perturbation 
theory, where ${\cal L} = H_0 - E  + V$, one chooses $P$ to project onto 
a particular eigenstate of $H_0$ and then the introduction of power series 
expansions into eqs.~(3.7) and (3.8) leads to familiar formulas.

We may illustrate this situation by choosing 
\bs
\begin{equation}
\mid \Psi\, \rangle = \; \mid p\, \rangle, \; \; \; \; \; \; (H_0 + V) \mid 
p\, \rangle = E_p \mid p\, \rangle, \; \; \; \; \; \; H_0 \mid p \, 
\rangle^0 = E^0_p \mid p \, \rangle^0 
\label{3.11a}
\end{equation}
\begin{equation}
\mid {\bf \cdot} \, \rangle = \; \mid p\, \rangle^0, \; \; \; \; \; \;
P = \; \mid p \, \rangle^{0 \; 0}\langle \, p \mid, \; \; \; \; \; \; 
\mid \Psi_P \, \rangle = \; \mid p \, \rangle^{0 \; 0}\langle \, p \mid 
p \, \rangle.   
\label{3.11b}
\end{equation}
\es
Equation (3.8) then becomes the energy equation,
\begin{equation}
E_p = E^0_p + V_{pp} + \sum_{m,n}\!^\prime \, V_{pm} [1 + Q(H_0-E_p)^{-1} 
QVQ]^{-1}_{mn} \frac{V_{np}}{(E_p-E^0_n)}   
\label{3.12}
\end{equation}
and equation (3.7) becomes the wavefunction equation,
\begin{equation}
^0\langle \, m\neq p \mid p \, \rangle = \{\sum_n \! ^\prime \, 
[1 + Q(H_0-E_p)^{-1} QVQ]^{-1}_{mn} \frac{V_{np}}{(E_p-E^0_n)}\}^0 
\langle \, p \mid p \, \rangle
\label{3.13}
\end{equation}
and both are easily iterated to any desired order of the perturbation $V$. 
In this example, each choice of projector $P$ is a choice to study a 
``nearby'' exact state $\mid \Psi \, \rangle$.

In the case of degenerate perturbation theory, one starts by choosing 
P as the projector into the degenerate subspace of interest and equation 
(3.8) becomes a matrix equation in that subspace.
 
For generalized Born-Oppenheimer problems, we proceed as follows.  
The original problem involves a number of coordinates, which we partition 
into two groups, called $x$ (the ``fast'' variables) and $y$ (the ``slow'' 
variables)
\begin{equation}
\mid \Psi\, \rangle = \; \mid \Psi(x,y)\, \rangle   
\label{3.14}
\end{equation}
and we choose a particular projector $P$ to act only on the $x$-variables,
\bs
\begin{equation}
\mid  {\bf \cdot} \, \rangle = \; \mid \psi_0(x)\, \rangle_R  
\label{3.15a}
\end{equation}
\begin{equation}
P = \; \mid  {\bf \cdot} \, \rangle \langle \, {\bf \cdot} \mid \; = \; 
\mid \psi_0(x) \, \rangle_R \int 
dx^\prime \: _R \langle \,\psi_0(x^\prime)\mid.
\label{3.15b}
\end{equation}
\es
In this class of applications, the partition into fast and slow variables 
and the choice of the projector state and its symmetry determines a 
preferred sector of the Hilbert space.  In practice, our choice of 
projector state $\mid  {\bf \cdot} \, \rangle$ below will be guided by the need to 
cancel the linear term in $R$ in (2.25).  The projected state is 
\bs
\begin{equation}
\mid \Psi_P\, \rangle = \; \mid \psi_0(x) \, \rangle_R \mid \psi(y) \, \rangle   
\label{3.16a}
\end{equation}
\begin{equation}
\mid \psi(y) \, \rangle = \langle \, {\bf \cdot} \mid \Psi(x,y) \, \rangle = 
\int dx^\prime \: _R \langle \,\psi_0(x^\prime) \mid \Psi(x^\prime,y) \, 
\rangle
\label{3.16b}
\end{equation}
\es   
where $\mid \psi(y)\, \rangle$ will be called the reduced state vector.  
Note that inner products with this projection operator involve integration 
over the fast variables x but not over the slow variables $y$; the symbol $R$ 
stands for a subset of the $y$ variables, and the subscript $R$ is placed on 
the projector state $\mid \psi_0(x)\, \rangle_R$ to indicate that this vector 
in the Hilbert space of the $x$-variables may be parametrized by some of the 
$y$-variables.

In these applications equations (3.4) and (3.8) are reduced Schrodinger 
equations in the slow variables $y$, the fast variables $x$ having being 
integrated out, and in particular eq.~(3.9)
\begin{equation}
\{\langle \, {\bf \cdot} \mid H-E \mid  {\bf \cdot} \, \rangle - \langle \, {\bf \cdot} 
\mid H Q (Q(H-E)Q)^{-1} Q H \mid  {\bf \cdot} \, \rangle \} \mid \psi(y)\, \rangle 
= 0
\label{3.17}
\end{equation}
is the effective Schrodinger equation for the reduced state vector 
$\mid \psi(y) \, \rangle$.  The terms of (3.17) are in 1-1 correspondence 
with the terms of (3.4), 
\begin{equation}
\int \! dx \, _R \langle \,\psi_0(x)\mid H-E \mid \psi_0(x) \, \rangle_R \mid 
\psi(y)\, \rangle + \int \! dx \, _R \langle \, \psi_0(x)\mid H \mid \Psi_Q(x,y)
\, \rangle = 0
\label{3.18}
\end{equation}
and the first term would give the ``first-order'' Born-Oppenheimer 
approximation (that is, eqs.~(2.24a,b)) if we were to ignore the second term.  
The second term can contribute in principle, however, to the effective 
Hamiltonian for $\mid \psi(y)\, \rangle$, and so we must proceed to solve the 
other equation (3.6)
\begin{equation}
Q H \mid \psi_0(x)\, \rangle_R \mid \psi(y)\, \rangle + Q (H-E) \mid 
\Psi_Q(x,y)\, \rangle = 0  
\label{3.19}
\end{equation}
for the state $\mid \Psi_Q(x,y)\, \rangle$.  If we have some small quantity, 
such as $\frac{1}{R}$ at large $R$, solutions of the system may be carried out 
in practice to any desired order of the small quantity.

Application of this machinery to matrix theory requires that we first 
make some transformations from the original variables.

\section{Canonical Transformations}

We focus now on the fermionic variables $\Lambda_{a\alpha}$ of matrix 
theory and carry out canonical transformations in order to introduce 
gauge-invariant fermions (Subsection 4.1) and to obtain a form of the 
Hamiltonian (Subsection 4.2) which is amenable to the computational method 
of the previous section. 

In what follows we scale out $\hbar$ and the coupling constant $g$, 
according to the relations
\bs
\begin{equation}
Q_\alpha(\hbar,g; \phi) = (g\hbar^2)^{\frac{1}{3} } Q_\alpha(1,1; 
\phi^\prime)    
\label{4.1a}
\end{equation}
\begin{equation}
H(\hbar,g; \phi) =  (g\hbar^2)^{\frac{2}{3}} H(1,1; \phi^\prime)    
\label{4.1b}
\end{equation}
\begin{equation}
\phi = (\frac{\hbar}{g})^{\frac{1}{3}} \phi^\prime 
\label{4.1c}
\end{equation}
\es
and it is really $\phi^\prime$ which appears below, although we drop the 
prime.  At any point, the reader may reinstate these parameters with the 
substitution
\begin{equation}
\phi \rightarrow (\frac{g}{\hbar})^{\frac{1}{3} } \phi  
\label{4.2}
\end{equation}
and the rescalings of $Q_\alpha$ and $H$ above.

\subsection{Gauge-Invariant Fermions}

Our first step involves the introduction of gauge-invariant fermions, 
using the eigenvectors $\psi_a^i$ which were introduced in (2.2).  The 
gauge-invariant fermions are defined as
\begin{equation}
\Lambda^\prime_{i\alpha} \equiv \psi_a^i \Lambda_{a\alpha}, \; \; \; \; \; \; i 
= 1,2,3, 
\; \; \; \; \; \; \alpha = 1 \ldots 16  
\label{4.3}
\end{equation}
and these preserve the anti-commutation relations 
\begin{equation}
\{\Lambda^\prime_{i\alpha}, \Lambda^\prime_{j\beta} \} = \delta_{ij} 
\delta_{\alpha\beta}.   
\label{4.4}
\end{equation}
Moreover, the gauge-invariant fermions allow us to write the Yukawa term 
in the Hamiltonian (1.5) as
\begin{equation}
H_F = -\frac{i}{2} \epsilon_{ijk} \Lambda^\prime_{i\alpha} 
(\Gamma_j)_{\alpha\beta} 
\Lambda^\prime_{k\beta} \lambda_j \equiv -\frac{i}{2} \epsilon_{ijk} 
(\Lambda^\prime_i 
\Gamma_j \Lambda^\prime_k) \lambda_j.
\label{4.5}
\end{equation}
The real symmetric, traceless and gauge-invariant matrices $\Gamma_i$ in 
(4.5) are defined by
\bs
\begin{equation}
\Gamma_i \equiv \frac{\Gamma^m  \phi^m_a  \psi_a^i}{\lambda_i} = \Gamma^m 
\eta^m_i,  
\; \; \; \; \; \; i = 1,2,3
\label{4.6a}
\end{equation}
\begin{equation}
\{\Gamma_i, \Gamma_j\} = 2\delta_{ij}
\label{4.6b}
\end{equation}
\es
and they preserve the Clifford algebra as shown.  (The 21 gauge-invariant 
angles $\eta^m_i$ are defined in (2.3).)  In what follows, $\Gamma_1 , 
\Gamma_2$ and $\Gamma_3$ are the components of $\Gamma_i$.

The eigenvectors $\psi_a^i$ are functions of the bosonic variables 
$\phi^m_a$, so the gauge-invariant fermions $\Lambda^\prime$ are 
coordinate-dependent and do not commute with the bosonic derivatives $\pi$.
We rectify this situation by making an additional canonical transformation 
to obtain independent bosonic momenta $\pi^\prime$:
\bs
\begin{equation}
\pi^{\prime m}_a = \pi^m_a + F^m_a
\label{4.7a}
\end{equation}
\begin{equation}
F^m_a = \frac{i}{2} (\Lambda^\prime_i (T^m_a)_{ij} \Lambda^\prime_j), \; 
\; \; \; \; \; 
(T^m_a)_{ij} = \psi_b^i \partial^m_a \psi_b^j 
\label{4.7b}
\end{equation}
\begin{equation}
[\pi^{\prime m}_a, \Lambda^\prime_{i\alpha}] = 0
\label{4.7c}
\end{equation}
\es
where $\pi^\prime$ and $\phi$ remain canonical. This allows us to specify 
that
\bs
\begin{equation}
\pi^{\prime m}_a \mid \Lambda^\prime \, \rangle = 0
\label{4.8a}
\end{equation}
\begin{equation}
\pi^{\prime m}_a [f(\phi) \mid \Lambda^\prime \, \rangle] = -i (\partial^m_a 
f(\phi)) \mid 
\Lambda^\prime \, \rangle 
\label{4.8b}
\end{equation}
\es
where $\mid \Lambda^\prime \, \rangle$ is any state formed with the 
gauge-invariant fermions.  
In what follows, we describe this situation by writing 
\begin{equation}
\pi^{\prime m}_a = -i \partial^m_a,\; \; \; \; \; \; \partial^m_a 
\Lambda^\prime = 0. 
\label{4.9}
\end{equation}
Further details of this transformation are given in Appendix A, which 
notes that the matrices $T^m_a$ are divergence-free flat connections.  

Appendix A also shows that the gauge generators (1.2) become purely 
bosonic
\begin{equation}
G_a = \epsilon_{abc} (\phi^m_b \pi^m_c - \frac{i}{2} (\Lambda _b \Lambda 
_c)) = 
\epsilon_{abc} \phi^m_b \pi^{\prime m}_c    
\label{4.10}
\end{equation}
when written in terms of the independent canonical momenta $\pi^\prime$.  
This 
result confirms that $G_a$ commutes with $\Lambda^\prime$ and tells us 
that states 
formed with the $\Lambda^\prime$ fermions
\begin{equation}
G_a f(\lambda, \eta) \mid \Lambda^\prime \, \rangle = 0
\label{4.11}
\end{equation}
are gauge invariant, as expected, when the bosonic coefficient $f$ is 
separately gauge invariant.

The rotation generators (1.3) also maintain a simple form
\begin{equation}
J^{mn} = \pi^{\prime[m}_a \phi^{n]}_a - \frac{1}{2} (\Lambda^\prime_i 
\Sigma^{mn} \Lambda^\prime_i)
\label{4.12}
\end{equation}
when expressed in terms of the independent momenta $\pi^\prime$. This 
result shows 
that, because $\psi_a^i$ is rotation-invariant, the gauge-invariant 
fermions 
remain spinors under spin$(9)$.  

Other applications of the gauge-invariant fermions include the 
following: The supercharges $Q_\alpha$ and the Hamiltonian $H$ can be 
written entirely in terms of gauge-invariant quantities.  This gauge-invariant 
formulation of $SU(2)$ matrix theory is given in Appendix B and continued in
Appendix G.  Moreover,  the complete diagonalization of the Yukawa term
\begin{equation}
H_F = - \sum_{k,\nu} \mu_k a^+_{k\nu} a_{k\nu}   
\label{4.13}
\end{equation}
is discussed in Appendix C.

\subsection{Further Transformations}

For our consideration below of large $R = \lambda_3$ asymptotic behavior, 
it is convenient to make another transformation to simplify the leading term in 
$H_F$, 
\begin{equation}
i \Lambda^\prime_{1\alpha} (\Gamma_3)_{\alpha\beta} 
\Lambda^\prime_{2\beta} 
\lambda_3 = i (\Lambda^\prime_1 \Gamma_3 \Lambda^\prime_2) R.
\label{4.14}
\end{equation}
We further define
\bs
\begin{equation}
\Lambda^{\prime \prime}_1 = \Gamma_3 \Lambda^\prime_1, \; \; \; \; \; \;
\Lambda^{\prime \prime}_2 = \Lambda^\prime_2, \; \; \; \; \; \;
\Lambda^{\prime \prime}_3 = \Lambda^\prime_3
\label{4.15a}
\end{equation}
\begin{equation}
\{\Lambda^{\prime\prime}_{i\alpha}, \Lambda^{\prime\prime}_{j\beta} \} 
= \delta_{ij} \delta_{\alpha\beta}
\label{4.15b}
\end{equation}
\es
along with another canonical transformation,
\begin{equation}
\pi^{\prime \prime m}_a = \pi^{\prime m}_a + G^m_a, \; \; \; \; \; \; 
[\pi^{\prime \prime m}_a, \Lambda^{\prime \prime}_{i\alpha}] = 0
\label{4.16}
\end{equation}
for which $\pi^{\prime \prime}$ and $\phi$ remain canonical.  See Appendix 
A for further details.

The final form of the gauge generators is 
\begin{equation}
G_a = \epsilon_{abc} \phi^m_b \pi^{\prime \prime m}_c
\label{4.17}
\end{equation}
because $\Lambda^{\prime \prime}$ are also gauge-invariant fermions.  The 
rotation generators 
are now
\begin{equation}
J^{mn} = \pi^{\prime \prime [m}_a \phi^{n]}_a - \frac{1}{2} 
(\Lambda^{\prime \prime}_i \Sigma^{mn} \Lambda^{\prime \prime}_i)
\label{4.18}
\end{equation}
and it follows that the $\Lambda^{\prime \prime}$ fermions remain spinors 
under spin$(9)$.

The final form of the Hamiltonian that results from our canonical 
transformations is the following (we now drop all primes for simplicity):
\begin{equation}
H = H_B + H_F + H_S
\label{4.19}
\end{equation}
where
\bs
\begin{eqnarray}
H_B & = & \frac{1}{2} \pi^m_a \pi^m_a + V, \\
\label{4.20a}
H_F & = & i (\Lambda_1 \Lambda_2) R + i (\Lambda_2 \Gamma_1 \Lambda_3) 
\lambda_1 + i (\Lambda_3 \Gamma_2 \Gamma_3 \Lambda_1) \lambda_2, \\
\label{4.20b}
H_S & = & - (F^m_a + G^m_a)\pi^m_a +\frac{1}{2} F^m_a F^m_a + \frac{1}{2} 
G^m_a G^m_a 
+ F^m_a G^m_a. 
\label{4.20c}
\end{eqnarray}
\es
Here
\bs
\begin{eqnarray}
\pi^m_a & \! = \! & -i \partial^m_a = -i \partial/\partial\phi^m_a, \; \; \; \; 
\; \; \partial^m_a 
\Lambda_{i\alpha} = 0 
\label{4.21a} \\
F^m_a & \! = \! & i(\Lambda_1 \Gamma_3 \Lambda_2) (T^m_a)_{12} + i(\Lambda_1 
\Gamma_3 \Lambda_3) (T^m_a)_{13} + i(\Lambda_2\Lambda_3) (T^m_a)_{23} \; \;
\label{4.21b} \\
G^m_a & \! = \! & \frac{i}{2} (\Lambda_1 (\Gamma_3 \partial^m_a \Gamma_3) 
\Lambda_1) 
\label{4.21c}
\end{eqnarray}
\es
and the connection $T$ is defined in (4.7).  Note that $H_F$ in (4.20b) is 
the original Yukawa term, now written in terms of the gauge-invariant 
fermions, and the shift term $H_S$, which is quartic in the gauge-invariant 
fermions, is the result of our canonical transformations.

Our third and final step is to introduce gauge-invariant fermion 
creation and annihilation operators for $\Lambda_1$  and $\Lambda_2$:
\bs
\begin{equation}
\Lambda_{1\alpha} = \frac{(a_\alpha + a_\alpha^+)}{\sqrt 2}, \; \; \; \; \; \; 
\Lambda_{2\alpha} = \frac{(a_\alpha - a_\alpha^+)}{i\sqrt 2}.
\label{4.22a}
\end{equation}
\begin{equation}
\{a_\alpha, a^+_\beta \} = \delta_{\alpha\beta}
\end{equation}
\es
This gives the first term in $H_F$ as
\begin{equation}
i (\Lambda_1 \Lambda_2) R = R (\sum_\alpha a_\alpha^+ a_\alpha - 8) 
\label{4.23}
\end{equation}
and the gauge-invariant empty state $\mid 0 \, \rangle$, defined by
\begin{equation}
a_\alpha \mid 0 \, \rangle = 0   
\label{4.24}
\end{equation}
gives the lowest value $-8R$ for this operator.

The final form of the rotation generators is
\begin{equation}
J^{mn} = \pi^{[m}_a \phi^{n]}_a - a^+_\alpha (\Sigma^{mn})_{\alpha\beta} 
a_\beta - \frac{1}{2} (\Lambda_3 \Sigma^{mn} \Lambda_3)
\label{4.25}
\end{equation}  
so that the state $\mid 0 \, \rangle$ is invariant under rotations of the 
$\Lambda_1, 
\Lambda_2$ fermions.

\section{The First Computation} 

The Hamiltonian (4.19) acts in the Hilbert space of the following 75 variables:
\begin{quote}
27 bosonic variables $\phi^n_a$, which we have packaged into 3 gauge- 
and rotational-invariant ``lengths'' --  $\lambda_1, \lambda_2, 
\lambda_3$ -- and 24 remaining ``angles'' $\Omega$.
\vspace{6pt}
\newline
48 fermionic operators, where 32 have been packaged into the 
gauge-invariant annihilation and creation operators $a_\alpha$ and 
$a_\alpha^+$ and another 16 gauge-invariant fermions $\Lambda_{3\alpha}$.
\end{quote}

We begin the computation by choosing a partition into the fast variables,
\begin{equation}
\mbox{``$x$'' variables:} \; \lambda_1, \lambda_2, a_\alpha, a_\alpha^+, 
\Omega  
\label{5.1}
\end{equation}
and the slow variables,
\begin{equation}
\mbox{``$y$'' variables:} \; \lambda_3 = R, \Lambda_{3\alpha}   
\label{5.2}
\end{equation}    
although we will discuss a slightly different partition in Section 7.   

Next, we must choose a particular projection operator $P$ and its 
associated projector state $\mid  {\bf \cdot} \, \rangle$.  Our choice is
\bs
\begin{equation}
\mid  {\bf \cdot} \, \rangle = \; \mid \psi_0(x)\, \rangle_R = u_R(\lambda_1, \lambda_2) \mid 
0 \, \rangle   
\label{5.3a}
\end{equation}
\begin{equation}
\mid \Psi_P\, \rangle = \; \mid  {\bf \cdot} \, \rangle \mid \psi(R, 
\Lambda_3) \, 
\rangle = \; \mid \psi_0(x)\, \rangle_R \: \mid \psi(R, \Lambda_3)\, \rangle 
\label{5.3b}
\end{equation}
\es
where $u_R(\lambda_1, \lambda_2)$ is defined in equation (2.18) and  $\mid 0 
\, \rangle$ is the empty fermion state for $\Lambda_1$ and $\Lambda_2$ defined 
in (4.24).  This state $\mid  {\bf \cdot} \, \rangle$ is the gauge-invariant analogue of the 
approximate ground state introduced in Ref.~\cite{Dan}, and, as discussed by 
these authors, it will guarantee the desired cancellation of the term linear in 
$R$ in the effective Hamiltonian (2.25). 

This leaves us to study the reduced state vector $\mid \psi(R, 
\Lambda_3)\, \rangle$ in the 
``$y$'' variables, 
\begin{equation}
\mid \psi(R, \Lambda_3)\, \rangle \! = \! \langle \, {\bf \cdot} \mid \Psi (x,y) \, 
\rangle \! = \! \langle \, {\bf \cdot} \mid \Psi_P\, \rangle \! = \! \int \! 
d^2\lambda (d\Omega) \sigma u_R^\ast (\lambda_1,\lambda_2) \langle \, 0 \mid 
\Psi(x,y)\, \rangle
\label{5.4}
\end{equation}
where $\sigma$ is defined in (2.17e)  and the normalization integral is 
\begin{equation}
\int dR R^{10} \langle \, \psi(R, \Lambda_3) \mid \psi(R, \Lambda_3)\, \rangle 
\, < \infty.
\label{5.5}
\end{equation}
Consistent with earlier notation, we define
\begin{equation}
\langle \, {\bf \cdot} \mid A \mid {\bf \cdot} \, \rangle \equiv \int d^2\lambda (d\Omega) \sigma 
u_R^\ast (\lambda_1,\lambda_2) \langle \, 0 \mid A \mid 0 \, \rangle 
u_R(\lambda_1,\lambda_2) 
\label{5.6}
\end{equation}    
where $A$ is any operator which may depend upon both the $x$ and $y$ 
variables.  The result of this partial average is an operator that depends 
only upon the $y$ variables and their derivatives.

We must now go through all the terms in the Hamiltonian (4.19) and 
answer the following questions for each operator:
\begin{enumerate}
\item What fermionic selection rules apply with respect to the number 
operator $N_F = \Sigma_\alpha a_\alpha^+ a_\alpha$?
\vspace{-6pt}
\item What is the order of magnitude of the operators in powers of $R$?
Here, it is important to remember that $\lambda_1$ and $\lambda_2$ are of 
order $R^{-\frac{1}{2}}$ at large $R$.
\end{enumerate}
The details of this assessment are given in Appendix  D.  The results 
are given below, phrased in the language of ``matrix elements,'' $PHP$, 
$PHQ$, $QHP$ and $QHQ$, as these appear in the basic equations (3.4) and (3.6).  
For reference, the first of these equations reads:
\begin{equation}
P(H-E)P \mid \Psi_P \, \rangle + PHQ \mid \Psi_Q \, \rangle = 0
\label{5.7}
\end{equation}
or, equivalently,
\begin{equation}
\langle \, {\bf \cdot} \mid H-E \mid  {\bf \cdot} \, \rangle\mid \psi(R, \Lambda_3)\, 
\rangle + \! \int \! d^2\lambda (d\Omega) \sigma u_R^\ast(\lambda_1,\lambda_2) 
\langle \, 0 \mid H \mid \Psi_Q(x,y) \, \rangle = 0
\label{5.8}
\end{equation}
and we begin by evaluating the first term of this equation.  
  
Generically, $H$ is dominated by terms of $O(R)$.  However, for the 
``diagonal matrix element'' $PHP$, these leading terms cancel, as 
anticipated in the discussion of Section 2.3,  and we are left with 
the terms through $O(R^{-2})$:
\begin{equation}
\langle \, {\bf \cdot} \mid H-E \mid {\bf \cdot} \, \rangle = -\frac{1}{2} 
\frac{d^2}{dR^2} - \frac{5}{R} \frac{d}{dR} + \frac{12}{R^2} - E + \ldots
\label{5.9}
\end{equation}
which will act on the reduced state vector $\mid \psi(R, \Lambda_3)\, \rangle$.  
Here, the dots indicate higher order terms in $\frac{1}{R}$.  If we were 
to stop here, the asymptotic effective Hamiltonian would be
\begin{equation}
H^{(1)}_{eff} = -\frac{1}{2} \frac{d^2}{dR^2} - \frac{5}{R} \frac{d}{dR} + 
\frac{12}{R^2}
\label{5.10}
\end{equation}
which defines the full first-order Born-Oppenheimer approximation, now
including the fermions. Comparing with the earlier bosonic result (2.25) for 
$H_{eff}$, we see that the fermionic contributions have cancelled the term 
linear in $R$ and added the term $+\frac{12}{R^2}$, which comes from the 
$\frac{(F^2 + G^2)}{2}$ terms in (4.20c).  But we cannot stop here because 
there are other terms of order $\frac{1}{R^2}$ to be found from the second term 
of (5.8) and to evaluate this term we need $\mid \Psi_Q(x,y) \, \rangle$.

To solve for $\mid \Psi_Q(x,y) \, \rangle$ we turn to the other basic 
equation (3.6) which we write as follows:
\begin{equation}
Q(H-E)Q \mid \Psi_Q\, \rangle + QHP \mid \Psi_P\, \rangle = 0
\label{5.11}
\end{equation}
or, equivalently,
\begin{equation}
Q(H-E)\mid \Psi_Q(x,y)\, \rangle + QH [u_R(\lambda_1,\lambda_2) \mid 0 \, 
\rangle \mid \psi (R, \Lambda_3)\, \rangle] = 0.
\label{5.12}
\end{equation}
The formal solution of this equation is
\begin{equation}
\mid \Psi_Q \, \rangle = -(Q(H-E)Q)^{-1} QHP \mid \Psi_P \, \rangle.
\label{5.13}
\end{equation}
    
For the term $QHP$ in (5.11) (an ``off-diagonal matrix element'') the 
leading contribution is of order $R^{-\frac{1}{2}}$ and comes only from 
the second and third terms of $H_F$ in (4.20b):
\bs
\begin{equation}
QHP \mid \Psi_P\, \rangle = Q \, H [u_R(\lambda_1,\lambda_2) \mid 0 \, 
\rangle \mid \psi (R, \Lambda_3) \, \rangle] \hspace{1.5in}
\label{5.14a}
\end{equation}
\begin{equation}
\simeq \frac{-\lambda_1(a^+ \Gamma_1 \Lambda_3) + i\lambda_2 (\Lambda_3 
\Gamma_2 \Gamma_3 a^+)}{\sqrt{2}} u_R (\lambda_1, \lambda_2)  \mid 0 
\, \rangle \mid \psi (R, \Lambda_3) \, \rangle. 
\label{5.14b}
\end{equation}
\es
The projection operator $Q$ does not appear in this last expression since 
we can write $Q = 1-P$ and $P$ annihilates (5.14b) because $\langle \, 0 \mid 
a^+ \mid 0 \, \rangle = 0$.  This state has $N_F=1$ and so the first term of 
(5.11) also has $N_F=1$.  

The leading terms in $QHQ$ (the ``energy denominator'') will have the 
generic $O(R)$ behavior of $H$.  The $O(R^{-\frac{1}{2}})$ estimate holds 
for $PHQ$ in equation (5.7), the same as for $QHP$. Then we can see that
the formal expression
\begin{equation}
(PHQ) (Q(H-E)Q)^{-1} (QHP) \sim O(R^{-\frac{1}{2}} 
(R-E+O(R^{-\frac{1}{2}}))^{-1} R^{-\frac{1}{2}})  
\label{5.15}
\end{equation}
(substitute (5.13) into (5.7)) will contribute a term of order $\frac{1}{R^2}$ 
to the (second term of the) reduced wave equation (5.7) and we must determine 
its numerical coefficient.  For this purpose we need compute only the terms of 
$O(R)$ in $QHQ$.

From the details in Appendix D we find that four terms in $H$ contribute 
to $QHQ$ at order $R$ and these include differential operators as 
well as multiplicative operators in the bosonic variables.  This makes the 
explicit inversion of the operator $Q(H-E)Q$ a difficult problem, so we 
shall go back to equation (5.11) and solve it, at large $R$, as an 
inhomogeneous differential equation for $\mid \Psi_Q \, \rangle$. This 
procedure is closely related to an early technique \cite{Dal,Charlie} 
in atomic physics.

To solve equation (5.11) make the asymptotic ansatz,
\begin{eqnarray}
\mid \Psi_Q (x,y) \, \rangle & = & \frac{- f_1 (\lambda_1, \lambda_2) 
(a^+ \Gamma_1 \Lambda_3) + i f_2(\lambda_1, \lambda_2) (\Lambda_3 \Gamma_2 
\Gamma_3 a^+)}{\sqrt{2}} \nonumber \\
\nonumber \\
& & \hspace{.25in} \times \; u_R(\lambda_1, \lambda_2) \mid 0 \, \rangle \mid 
\psi (R, \Lambda_3) \, \rangle 
\label{5.16}
\end{eqnarray}
which is modeled after equation (5.14), with the insertion of two unknown 
functions $f_1$ and $f_2$. The ansatz is again annihilated by $P$ because 
\linebreak $\langle \, 0 \mid a^+ \mid 0 \, \rangle = 0$.  
Calculating the action of $Q(H-E)Q$ on this $\mid \Psi_Q\, \rangle$, we find 
(see Appendix D) that the asymptotic form of $Q(H-E) \mid 
\Psi_Q(x,y)\, \rangle$ has the same form as (5.14), involving the 
same fermion bilinears  $(a^+ \Gamma_1 \Lambda_3)$ and 
$(\Lambda_3 \Gamma_2 \Gamma_3 a^+)$.  Setting the total coefficients of 
these fermion bilinears to zero in (5.11) gives the coupled inhomogeneous 
differential equations
\bs
\begin{equation}
[-\frac{1}{2} (\Delta_1 + \Delta_2) + {\rm RD} + \frac{3}{\lambda_1^2} + 
{\rm U} + {\rm R} - {\rm E}] f_1 - {\rm Z} f_2 = -\lambda_1 
\label{5.17a} 
\end{equation}
\begin{equation}
[-\frac{1}{2} (\Delta_1 + \Delta_2) + {\rm RD} + \frac{3}{\lambda_2^2} + 
{\rm U} + {\rm R} 
- {\rm E}] f_2 - {\rm Z} f_1 = -\lambda_2 
\label{5.17b}
\end{equation}
\begin{eqnarray}
\Delta_1 & \equiv & (\frac{\partial}{\partial\lambda_1})^2 + 
[\frac{6}{\lambda_1} + 2 
\frac{\lambda_1}{(\lambda_1^2 - \lambda_2^2)}] 
\frac{\partial}{\partial\lambda_1} 
\label{5.17c} \\
\Delta_2 & \equiv & (\frac{\partial}{\partial\lambda_2})^2 + 
[\frac{6}{\lambda_2} + 2 
\frac{\lambda_2}{(\lambda_2^2 - \lambda_1^2)}] 
\frac{\partial}{\partial\lambda_2} 
\label{5.17d} \\
D & \equiv & \lambda_1 \frac{\partial}{\partial\lambda_1} + \lambda_2 
\frac{\partial}{\partial\lambda_2}
\label{5.17e} \\
U & \equiv & \frac{\lambda_1^2 + \lambda_2^2}{(\lambda_1^2 - \lambda_2^2)^2}
\label{5.17f} \\
Z & \equiv & \frac{2 \lambda_1 \lambda_2}{(\lambda_1^2 - \lambda_2^2)^2}
\label{5.17g}
\end{eqnarray}
\es
for the unknown functions $f_1$ and $f_2$.  As planned, we have kept only 
terms of $O(R)$ multiplying $f_1$ and $f_2$ on the left of these 
equations, 
and the inhomogeneous terms on the right come from the $\lambda_1$ and 
$\lambda_2$ factors in (5.14b).

In fact, we have found a simple exact particular solution of these 
equations:
\begin{equation}
f_1 = -\frac{\lambda_1}{(2R-E)},  \;\;\;   f_2 = -\frac{\lambda_2}{(2R-E)} 
.
\label{5.18}
\end{equation}
(Barring such luck, we would have carried out numerical computations, 
using, for example, the variational principle mentioned earlier.)   

The general solution to (5.17) can also include any solution to the 
homogeneous version of the equations, in addition to this particular 
solution.  Because of the non-vanishing linear terms in $R$ in these 
equations (which represent $Q(H-E)Q$), any solutions of the homogeneous 
equations will decay exponentially at large $R$, as in equation (2.27),  and 
can thus be consistently ignored compared to the asymptotic power-law 
behavior expected for the reduced state vector $\mid \psi(R, 
\Lambda_3)\, \rangle$.

Now that we know $\mid \Psi_Q(x,y)\, \rangle$ in (5.16), we can compute the 
large $R$ contribution to the second term of (5.7):
\bs
\begin{eqnarray}
\langle \, {\bf \cdot} \mid HQ \mid \Psi_Q\, \rangle & = & \langle \, {\bf \cdot} 
\mid \frac{\lambda_1(a \Gamma_1 \Lambda_3) + i\lambda_2 (\Lambda_3 
\Gamma_2 \Gamma_3 a)}{\sqrt{2}} \mid \Psi_Q(x,y) \, \rangle \; \; \; \; 
\; \; \; \; \; \; \; \; \;
\label{5.19a} \\
& = & - \, \frac{1}{(2(2R-E))} \langle \, {\bf \cdot} \mid(\Lambda_3 
(\lambda_1^2 + \lambda_2^2 + 2 \lambda_1 \lambda_2 \Theta)\Lambda_3)\mid  
{\bf \cdot} \, \rangle \nonumber \\
\nonumber \\
& & \hspace{.25in} \times \mid \psi(R, \Lambda_3)\, \rangle 
\label{5.19b} \\
& = & - \, \frac{1}{(2(2R-E))} \int d^2\lambda (d\Omega) \sigma \mid 
u_R(\lambda_1, \lambda_2)\mid^2 \nonumber \\
\nonumber \\ 
& & \hspace{.25in} \times \, (\Lambda_3 (\lambda_1^2 + \lambda_2^2 + 2
\lambda_1 \lambda_2 \Theta)\Lambda_3) \mid \psi(R, \Lambda_3)\, \rangle.
\label{5.19c}
\end{eqnarray}
\es
The gauge-invariant matrix $\Theta$ in (5.19) is defined as
\begin{equation}
\Theta \equiv -i  \Gamma_1 \Gamma_2 \Gamma_3
\label{5.20}
\end{equation}
and when we take the average over angles, 
\begin{equation}
\int  (d\Omega) \Theta = 0
\label{5.21}
\end{equation}
because $\Theta$ is odd under reflection of all the $\phi$ variables (see 
(E.18)).  From Appendix F we find that the average value of 
$(\lambda_1^2 + \lambda_2^2)$ is $\frac{8}{R}$; and from the anticommutation 
relations we have $(\Lambda_3 \Lambda_3) = 8$, so that our result is 
independent of any representation we might choose for the 
$\Lambda_3$ variables.  Then the result of this ``second order'' 
calculation, 
\begin{equation}
\langle \,{\bf \cdot} \mid HQ \mid \Psi_Q\, \rangle = -\frac{16}{R^2} \mid 
\psi(R, \Lambda_3)\, \rangle
\label{5.22}
\end{equation}
is exact through order $\frac{1}{R^2}$.

\section{The First Set of Candidate Ground States}  

Adding the result in eq.~(5.22) to the ``first order'' terms in eq.~(5.10), 
we find the asymptotic form of eq.~(5.7), exact through order $\frac{1}{R^2}$:
\bs
\begin{eqnarray}
H_{eff} \mid \psi(R, \Lambda_3)\, \rangle & = & E \mid \psi(R, 
\Lambda_3)\, \rangle
\label{6.1a} \\
H_{eff} & = & -\frac{1}{2} \frac{d^2}{dR^2} - \frac{5}{R} \frac{d}{dR} - 
\frac{4}{R^2}.
\label{6.1b}
\end{eqnarray}
\es
This reduced Schrodinger equation has two solutions at $E=0$: $R^{-1}$ or 
$R^{-8}$ times any state $\mid \Lambda_3 \, \rangle$ formed with the 
gauge-invariant fermions $\Lambda_3$.  The second solution
\begin{equation}
\mid \psi(R, \Lambda_3)\, \rangle  \simeq  R^{-8} \mid\Lambda_3\, \rangle  
\label{6.2}
\end{equation}
allows the normalization integral (5.5) to converge at large $R$.  (In the 
language of eq.~(2.28), we have found that $B^\prime = 8 < \frac{77}{4}$.)  
The result (6.2) is our first asymptotic set of ground state candidates, which 
must be tested further for global stability at non-asymptotic values of $R$.

These solutions also confirm \cite{Lusch} a continuous spectrum for $E > 0$.  
With $E = \frac{k^2}{2}$, the effective Hamiltonian (6.1) yields plane-wave 
normalizable solutions which behave as 
\begin{equation}
\mid \psi(R, \Lambda_3)\, \rangle_\pm  \simeq  R^{-5} e^{\pm ikR} 
\mid\Lambda_3\, \rangle
\label{6.3}
\end{equation}
at large $R$.

We can also follow the computation backward to reconstruct the asymptotic 
form of the candidate ground states $\mid \Psi\, \rangle$ near the flat 
directions of the potential $V$.  Using eqs.~(5.3), (5.16) and (6.2), we find 
the asymptotic forms
\bs
\begin{eqnarray}
\mid \Psi_P\, \rangle & = & R^{-8} u_R(\lambda_1, \lambda_2)  \mid 0 \, \rangle 
\mid \Lambda_3 \, \rangle
\label{6.4a} \\
\mid \Psi_Q\, \rangle & = & \frac{1}{(2\sqrt{2} R)} [\lambda_1 (a^+ \Gamma_1 
\Lambda_3) - i \lambda_2 (\Lambda_3 \Gamma_2 \Gamma_3 a^+)] \mid \Psi_P\, 
\rangle 
\label{6.4b}
\end{eqnarray}
\es
where $\mid 0 \, \rangle$ is the ground state of the gauge-invariant fermions 
$\Lambda_{1,2}$ and $u_R$ is given in eq.~(2.18).

Adding these results, we obtain the full asymptotic form of the candidate 
ground states, up to an overall normalization constant,
\bs
\begin{eqnarray}
\mid \Psi\, \rangle & \simeq & \{1 + \frac{R^{-\frac{3}{2}}}{(2\sqrt{2})} [z_1 
(a^+ \Gamma_1 \Lambda_3) - i z_2 (\Lambda_3 \Gamma_2 \Gamma_3 a^+)] \} 
\nonumber \\
& & \hspace{.25in} \times R^{-4} \: \exp (-\frac{(z_1^2 + z_1^2)}{2}) \mid 0 \, \rangle \mid 
\Lambda_3\, \rangle
\label{6.5a} \\
z_1 & = & \lambda_1 R^{\frac{1}{2}}, \; \; \; \; \; \;  z_2=  \lambda_2 
R^{\frac{1}{2}}
\label{6.5b}
\end{eqnarray}
\es
where the scaled variables $z_1$ and $z_2$ are those defined earlier in 
eq.~(2.22).  In this form of the candidate ground states, the variables
$z_{1,2}$ are finite and only R is large.

This result can also be written through this order in $\frac{1}{R}$ as
\bs
\begin{eqnarray}
\mid \Psi\, \rangle & \simeq & \exp (-\frac{S}{\hbar}) \mid 0 \, \rangle \mid 
\Lambda_3 \, \rangle
\label{6.6a} \\
S & = & \frac{V}{gr} + \{\frac{H_F}{2gr} + 4 \hbar \ln r\}
\label{6.6b} \\
r & = & (\lambda_1^2 + \lambda_2^2 + \lambda_3^2)^{\frac{1}{2}} = 
(\phi^m_a \phi^m_a)^{\frac{1}{2}} = R + O(R^{-2})
\label{6.6c} \\
\lambda_{1,2} & = & O ((\frac{\hbar}{g})^{\frac{1}{2}} R^{-\frac{1}{2}}) = 
O ((\frac{\hbar}{g})^{\frac{1}{2}} r^{-\frac{1}{2}})
\label{6.6d}
\end{eqnarray}
\es
where $V$ is the bosonic potential, $H_F$ is the Yukawa term and we have 
reinstated $\hbar$ and $g$ following the rule (4.2).  The range of validity 
in (6.6d) (finite $z_{1,2}$ at large r) defines the quantum neighborhood of the 
classical flat direction 
$\lambda_1 = \lambda_2 = 0$.  This form of the result shows that these candidate 
ground states are quantum extensions of Itoyama's solution of the zero-energy 
Hamilton-Jacobi equation, now made normalizable by the quantum correction 
$4 \hbar \ln r$ (this corresponds to $\alpha = 4 > \frac{3}{2}$ in the 
discussion of Section 2.2).

This set of candidate ground states does not include a singlet under 
spin$(9)$.  To see this explicitly, we note that the bosonic prefactor   
$\exp(-\frac{S}{\hbar})$ in (6.6) is rotation invariant while the rotation 
generators (4.25) give
\bs
\begin{eqnarray}
J^{mn} \mid 0 \, \rangle \mid \Lambda_3 \, \rangle & = & \mid 0 \, \rangle 
(-\frac{1}{2} (\Lambda_3 \Sigma^{mn} \Lambda_3)) \mid \Lambda_3 \, \rangle
\label{6.7a} \\
\frac{1}{2} J^{mn} J^{mn} \mid 0 \, \rangle \mid \Lambda_3 \, \rangle & = & 18  
\mid 0 \, \rangle \mid 
\Lambda_3 \, \rangle
\label{6.7b}
\end{eqnarray}
\es
on the fermion states.  The evaluation of the Casimir operator in (6.7b) follows 
from Fierz transformations and properties of the $\Gamma$ matrices.  This shows 
that all three irreducible representations of spin$(9)$ in $\mid \Lambda_3 
\, \rangle$ (and hence in the candidate ground states)
\begin{equation}
\mid \Lambda_3 \, \rangle \; \; \; = \; \; \; \mid 
256\, \rangle \; \; \; = \; \; \;  
\mid 44\, \rangle \oplus  \mid 84\, \rangle  \oplus  \mid 128\, \rangle  
\label{6.8}
\end{equation}
have the same value of the Casimir.  These irreps correspond respectively 
to the spin$(9)$ irreps of the 11-dimensional supergraviton:
\begin{enumerate}
\item a symmetric, traceless second rank tensor $(g_{mn})$
\vspace{-6pt}
\item a totally antisymmetric third rank tensor $(H_{mnp})$
\vspace{-6pt}
\item a ``gravitino'' or Rarita-Schwinger irrep $(B^m_\alpha)$
\end{enumerate}
where the first two irreps are bosonic and the last is fermionic.

We also remark that the set (6.6) of $256$ candidate ground states forms 
a ``zero-index unit'' whose presence cannot violate the index theorem 
\cite{Seth} for $SU(2)$ matrix theory.  In this connection, it is clear 
that there are strong parallels between our generalized Born-Oppenheimer 
formulation and the computational method of Ref.~\cite{Seth}. It is difficult 
to make a quantitative comparison, however, because we are computing different
quantities.

\section {A More General Set of Candidates} 

Having obtained our first set (6.6) of ground state candidates, we are 
now in a position to obtain a more general set of candidates.  

One crucial observation is that the angular integration $(d\Omega)$ played 
a very limited role in the computation of Section 5: Because our projector state 
$\mid {\bf \cdot} \, \rangle$ was independent of the angular variables $\Omega$, 
we needed only $\int(d\Omega) =1$ in every stage except for eq.~(5.19), where 
$\int(d\Omega) \Theta = 0$ eliminated the term proportional to the operator 
$\Theta$.  This opens the possibility of broadening our perspective by 
partitioning the variables $\Omega$ into fast and slow variables, while 
maintaining the requirement that no linear terms in $R$ should appear in the 
effective Hamiltonian.  In what follows, we ignore the 3 gauge degrees of
freedom in $\Omega$, keeping only the 21 gauge-invariant angular variables 
$\eta$ which we give again here for reference,
\begin{equation}
\eta^m_i =  \frac{\phi^m_a \psi_a^i}{\lambda_i}, \; \; \; \; \; \; 
\eta^m_i \eta^m_j = \delta_{ij}.    
\label{7.1}
\end{equation}
Recall that these 21 variables plus the 3 $\lambda$'s are a complete set of 
$27-3=24$ gauge-invariant bosonic variables for $SU(2)$.

More precisely, we begin our second computation by choosing the partition
\bs
\begin{eqnarray}
\mbox{fast $(x)$ variables:} & \: & \lambda_1, \lambda_2, \Lambda_1, 
\Lambda_2, \eta_1, \eta_2
\label{7.2a} \\
\mbox{slow $(y)$ variables:} & \: & \lambda_3, \Lambda_3, \eta_3
\label{7.2b}
\end{eqnarray}
\es
because, as demonstrated below, this will allow us to avoid $R$ terms in 
the effective Hamiltonian for the reduced state vector $\mid \psi(R, 
\Lambda_3, \eta_3) \, \rangle$.  Moreover, we choose the same $\eta_1, 
\eta_2$-independent projector state 
\begin{equation}
\mid  {\bf \cdot} \, \rangle =  u_R (\lambda_1, \lambda_2) \mid 0 \, \rangle
\label{7.3}
\end{equation}
used in the first computation, but now we must specify the decomposition 
of the $\eta$-measure in order to integrate out the fast variables 
$\eta_1$ and $\eta_2$.  The full gauge-invariant measure can be written
\bs
\begin{eqnarray}
(d\phi) & = & d^3\lambda \rho(\lambda) (d^3\eta)
\label{7.4a} \\
(d^3\eta) & = & (d^2\eta) (d\eta_3) 
\label{7.4b} \\
(d^2\eta) & = & [\prod_{i=1,2} (\prod_{m=1}^9 d\eta^m_i) 
\delta(\eta^n_i \eta^n_i -1) \delta(\eta^p_i \eta^p_3)] \delta(\eta^q_1 
\eta^q_2)
\label{7.4c} \\
(d\eta_3) & = & (\prod_{m=1}^9 d \eta^m_3) \delta(\eta^n_3 \eta^n_3 -1)
\label{7.4d}
\end{eqnarray}
\es
and, in what follows, we will need only the following two properties of 
$(d^2\eta)$,
\bs
\begin{eqnarray}
\int(d^2\eta) & = 1
\label{7.5a} \\
\int(d^2\eta) \Theta & = & 0.
\label{7.5b} 
\end{eqnarray}
\es
It is straightforward to see that $\int(d^2\eta)$ is independent of 
$\eta_3$ and (7.5a) is a convenient convention.  The property in (7.5b) 
follows because the matrix $\Theta$
\begin{equation}
\Theta = -i \Gamma_1 \Gamma_2 \Gamma_3 = -i \Gamma^m \Gamma^n \Gamma^p \eta^m_1 
\eta^n_2 \eta^p_3   
\label{7.6}
\end{equation}
is odd in each of the three $\eta$'s while $(d^2\eta)$ is even in $\eta_1$ 
and/or
$\eta_2$.

Relative to our first computation, we now have the replacement
\begin{equation}
(d\Omega) \rightarrow (d^2\eta) 
\label{7.7}
\end{equation} 
in all averages over fast variables.  This includes, for example, the new 
form of eq.~(5.6)
\begin{equation}
\langle \, {\bf \cdot} \mid A \mid  {\bf \cdot} \, \rangle =  \int d^2 \lambda (d^2\eta) \sigma 
u_R^\ast (\lambda_1, 
\lambda_2) \langle \, 0\mid A \mid 0\, \rangle u_R (\lambda_1, \lambda_2).
\label{7.8}
\end{equation}
Correspondingly, the integration over $\eta_3$ appears only in the new 
normalization condition
\begin{equation}
\int dR R^{10} (d\eta_3) \langle \, \psi(R, \Lambda_3, \eta_3) \mid \psi(R, 
\Lambda_3, \eta_3) \; \, \rangle  \; < \; \infty
\label{7.9}
\end{equation} 
for the reduced state vector.

The second computation (see Appendix D) then proceeds exactly as did 
the first computation, using the same ansatz (5.16) for $\mid \Psi_Q 
\, \rangle$ now with
\begin{equation}
\mid \psi (R,\Lambda_3) \, \rangle \rightarrow \mid \psi (R,\Lambda_3, 
\eta_3)\, \rangle  
\label{7.10}
\end{equation}
for the reduced state vector.  The same contributions are obtained (now by 
$\int(d^2\eta) = 1$) from each term, including the elimination of the $\Theta$ 
term (now by $\int(d^2\eta) \Theta = 0$) in the new version of eq.~(5.19).  
There is, however, one new contribution to $PHP$ from the action of the Laplacian 
$\Delta$ on the angular variables $\eta_3$ of the reduced state vector
$\mid \psi (R, \Lambda_3, \eta_3) \, \rangle$.  We sketch here only the 
asymptotic results that we need for this computation, referring the reader 
to Appendix G for the full structure of the Laplacian on general gauge-invariant 
functions $f(\lambda,\eta)$.

To study the new contribution of the Laplacian, we begin with the identity
\begin{equation}
(\partial^m_a \eta^n_i) (\partial^m_a \lambda_j) = 0
\label{7.11}
\end{equation}
which is, in fact, equivalent to eq.~(E.14).  It follows that the 
Laplacian is separable in the form
\begin{equation}
\Delta = \Delta_\lambda + \Delta_\eta
\label{7.12}
\end{equation}
where $\Delta_\lambda$, which contains the $\lambda$ derivatives, is 
defined in (2.4b) and $\Delta_\eta$ contains only derivatives with respect to the 
$\eta$ variables.  With the chain rule and the asymptotic identity
\begin{equation}
\partial^m_a \eta^n_3 = \frac{1}{R} (\delta^{mn} - \eta^m_3 \eta^n_3) 
\psi_a^3 + 
O(R^{-\frac{5}{2}})
\label{7.13}
\end{equation}
we can easily compute the extra asymptotic contribution to $PHP$ as
\bs
\begin{eqnarray}
& & \! \langle \, {\bf \cdot} \mid -\frac{1}{2} \Delta_\eta \mid  {\bf \cdot} \, \rangle \mid \psi (R, 
\Lambda_3, \eta_3) \, \rangle
\label{7.14a} \\
& = & \! \int d^2 \lambda (d^2\eta) \sigma u_R^\ast (\lambda_1, \lambda_2) 
(-\frac{1}{2} \Delta_\eta) u_R (\lambda_1, \lambda_2) \mid \psi (R, 
\Lambda_3, \eta_3) \, \rangle 
\label{7.14b} \\
& = & \! (\frac{L_3^2}{2R^2} + O(R^{-3})) \mid \psi(R, \Lambda_3, \eta_3) 
\, \rangle.
\label{7.14c}
\end{eqnarray}
\es
Derivatives with respect to $\eta_1$ and $\eta_2$ do not contribute in 
this 
computation because there is no dependence on these variables in 
$u_R(\lambda_1, 
\lambda_2)$ or the reduced state vector $\mid \psi(R, \Lambda_3, 
\eta_3)\, \rangle$.  
Moreover, we have organized the result into the angular momentum operators 
of 
$\eta_3$,
\begin{equation}
L^{mn}_3 = - i \eta ^{[m}_3 \partial^{n]}_3, \; \; \; \; \; \; \; L_3^2 = 
\frac{1}{2} L^{mn}_3 L^{mn}_3
\label{7.15}
\end{equation}
where $\partial^m_3 = \frac{\partial}{\partial\eta^m_3}$.  In this result, 
the $\eta$ derivatives may be taken as the naive derivative
\begin{equation}
\partial^m_3 \eta^n_3 = \delta^{mn}
\label{7.16}
\end{equation}
or the constrained derivative
\begin{equation}
\partial^m_3 \eta^n_3 = \delta^{mn} - \eta^m_3 \eta^n_3 
\label{7.17}
\end{equation}
which respects the constraint $\eta^m_3 \eta^m_3 = 1$:  Both give the same 
operators 
$L^{mn}_3$, which generate a bosonic $SO(9)$.   

Adding this extra term then, we find the new asymptotic effective 
Hamiltonian
\bs
\begin{equation}
H_{eff}  \mid \psi(R, \Lambda_3, \eta_3) \, \rangle = E \mid \psi(R, 
\Lambda_3, \eta_3)\, \rangle 
\label{7.18a}
\end{equation}
\begin{equation}
H_{eff} = -\frac{1}{2} \frac{d^2}{dR^2} - \frac{5}{R} \frac{d}{dR} + 
\frac{L_3^2 - 8}{2R^2}
\label{7.18b}
\end{equation}
\es
which is exact through $O(R^{-2})$.  This reduced system becomes a simple 
radial wave equation when we introduce the spherical harmonics 
$Y_l(\eta_3)$
of $SO(9)$
\begin{equation}
L_3^2 Y_l(\eta_3) = l(l+7) Y_l(\eta_3), \; \; \; \; \; \; \;  l= 0,1,2 
\ldots
\label{7.19}
\end{equation}
with ``magnetic'' degeneracy 
\begin{equation}
\deg(l) = \frac{(2l+7) (l+6)!}{l ! 7 !}.
\label{7.20}
\end{equation}
Using the spherical harmonics, we can immediately write down the 
normalizable asymptotic solutions
\bs
\begin{equation}
\mid \psi(R, \Lambda_3, \eta_3) \, \rangle \simeq  R^{-(l+8)} Y_l(\eta_3) 
\mid\Lambda_3 \, \rangle
\label{7.21a}
\end{equation}
\begin{equation}
\mid \Lambda_3 \, \rangle  = \, \mid 256 \, \rangle = \, \mid 44 \, \rangle 
\, \oplus \mid 84 \, \rangle \, \oplus \mid 128 \, \rangle
\label{7.21b}
\end{equation}
\es
for the reduced state vector.

This set of solutions contains our first solution (6.2) as the special 
case with $l = 0$, and the set contains exactly one state
\begin{equation}
\mid \psi(R, \Lambda_3, \eta_3) \, \rangle_{l=2} \; \simeq R^{-10} Y_2^{mn} 
(\eta_3) \mid 44; mn \, \rangle
\label{7.22}
\end{equation}
which is a singlet under spin$(9)$.  Here we have used an explicit form of 
the 44-dimensional $Y_2(\eta_3)$ 
\begin{equation}
Y_2^{mn}(\eta_3) = \eta^m_3 \eta^n_3 - \frac{1}{9} \delta^{mn}
\label{7.23}
\end{equation}
to perform the invariant sum over $Y_2$ times the 44-dimensional irrep in 
$\mid \Lambda_3\, \rangle$.

The new effective Hamiltonian (7.18) also exhibits plane-wave normalizable 
solutions
\begin{equation}
\mid \psi(R, \Lambda_3, \eta_3) \, \rangle_\pm \: \simeq \frac{Y_l(\eta_3) e^{\pm 
ikR}}{R^5} \mid \Lambda_3 \, \rangle  
\label{7.24}
\end{equation}
and hence a continuous spectrum for $E = \frac{k^2}{2} > 0$.  The 
earlier result (6.3) is included in (7.24) when $l=0$.

Finally, we may follow the new computation backward to obtain the full 
asymptotic form of our general set of candidate $SUSY$ ground states.  One 
obtains the generalization of (6.5),
\bs
\begin{eqnarray}
\mid \Psi\, \rangle & \simeq & \{1 + \frac{R^{-\frac{3}{2}}}{(2\sqrt{2})} [z_1 
(a^+ \Gamma_1 \Lambda_3) - i z_2 (\Lambda_3 \Gamma_2 \Gamma_3 a^+)] \} 
\nonumber \\
& & \hspace{.25in} \times R^{-l-4} Y_l(\eta_3) \: \exp (-\frac{(z_1^2 + z_1^2)}
{2}) \mid 0 \, \rangle \mid \Lambda_3\, \rangle
\label{7.25a} \\
z_1 & = & \lambda_1 R^{\frac{1}{2}}, \; \; \; \; \; \;  z_2=  \lambda_2 
R^{\frac{1}{2}}
\label{7.25b}
\end{eqnarray}
\es
and the generalization of (6.6),
\bs
\begin{eqnarray}
\mid \Psi \, \rangle & \simeq & \exp (-\frac{S}{\hbar}) Y_l(\eta_3) \mid  0 
\, \rangle \mid  \Lambda_3 \, \rangle
\label{7.26a} \\
S & = & \frac{V}{gr} + \{\frac{H_F}{2gr} + (l+4) \hbar \ln r\}
\label{7.26b} \\
r & = & (\lambda_1^2 + \lambda_2^2 + \lambda_3^2)^{\frac{1}{2}} = 
(\phi^m_a \phi^m_a)^{\frac{1}{2}} 
\label{7.26c} \\
\lambda_{1,2} & = & O ((\frac{\hbar}{g})^{\frac{1}{2}} r^{-\frac{1}{2}})
\label{7.26d} \\
\mid \Psi\, \rangle_{l=2} & \simeq & \exp (-\frac{S}{\hbar}) Y_2^{mn}(\eta_3) 
\mid  0 \, \rangle \mid  44; mn \, \rangle
\label{7.26e} \\
J^{mn} \mid \Psi \, \rangle_{l=2} & = & 0
\label{7.26f}
\end{eqnarray}
\es
where we have recorded in (7.26e) the unique candidate which is a singlet 
under spin$(9)$.  The extreme semiclassical limit of the $l\neq 0$ solutions in 
(7.26) are gauge- but not rotation-invariant solutions of the zero-energy 
Hamilton-Jacobi equation.

The apparent simplicity of the ground state candidates (7.26a,b) 
suggests that there may be a more elegant path to this result.

Each of these candidates is a normalizable zero-energy asymptotic 
solution of the Schrodinger equation of $SU(2)$ matrix theory, but 
each must be tested further for stability at non-asymptotic values of $R$.  The 
high-$l$ solutions are particularly suspect because they are associated to 
the growing centrifugal barrier $\frac{L_3^2}{2R^2}$.  Since the singlet 
state has $l=2$, this leaves the states with $l \leq 2$ as the most auspicious 
candidates.

Appendix G also outlines a strategy for a proof of a conjecture 
which, if true, would tell us that our projector state in (7.3) is the only 
state in the Hilbert space of the fast variables (7.2a) that avoids 
linear terms in $R$ in the effective Hamiltonian. In this case, our set of 
candidate ground states would be a complete list for the partition (7.2) of
the variables of $SU(2)$ matrix theory.

\vspace{1.5in}

\noindent {\bf ACKNOWLEDGEMENTS}

      For helpful discussions, we thank J. de Boer, H. Itoyama, 
H.Murayama, H. Nicolai, H. Ooguri,  P. Pouliot and P. Yi. 

     The work of  M.B.H. was supported in part by the Director, Office of 
Energy Research, Office of Basic Energy Sciences, of the U.S. Department 
of Energy under Contract DE-AC03-76F00098 and in part by the National 
Science Foundation under grant PHY95-14797.

\vskip 1.0cm
\setcounter{equation}{0}
\def\theequation{A.\arabic{equation}}
\boldmath
\noindent{\bf Appendix A: Canonical Transformations}
\unboldmath
\vskip 0.5cm

In further detail, the shift in (4.7) is
\bs
\begin{eqnarray}
\pi^{^\prime m }_a & = & \pi^m_a + F^m_a 
\label {A1a} \\
F^m_a & = & \frac{1}{2i} \Lambda_{b \alpha} (S^m_a)_{bc} 
\Lambda_{c\alpha}= \frac{i}{2}  \Lambda^\prime_{i\alpha} (T^m_a)_{ij} 
\Lambda^\prime_{j\alpha}
\label{A1b} \\ 
(S^m_a)_{bc} & = & \psi_b^i \partial^m_a \psi_c^i, \; \; \; \; \; 
(T^m_a)_{ij}= \psi_b^i \partial^m_a \psi_b^j 
\label{A1c}
\end{eqnarray}
\es
where $\Lambda^\prime$ are the gauge-invariant fermions (4.3) and 
$\psi_a^i$ are the eigenvectors of $\Phi$ introduced in (2.2).  Further 
properties of $T^m_a$ are found in Appendix E.  The quantities $S^m_a$ 
and $T^m_a$ are real antisymmetric matrices, which we call connections.

Using the orthonormality and completeness of $\psi_a^i$ in (2.2), one 
verifies that $\phi$ and $\pi^\prime$ are canonical variables which are 
independent of $\Lambda^\prime$:
\bs
\begin{eqnarray}
\left[ \pi^{\prime m}_a, \Lambda^\prime_{i\alpha}\right] & = & 0
\label{A2a} \\
\left[ \phi^m_a, \pi^{\prime n}_b \right] & = & i \delta_{mn} \delta_{ab}
\label{A2b} \\
\left[ \pi^{\prime m}_a, \pi^{\prime n}_b \right] & = & 0
\label{A2c}
\end{eqnarray}
\es
and this tells us that 
\begin{equation}
\pi^{\prime m} _a = -i \partial^m_a = -i 
\frac{\partial}{\partial\phi^m_a}, 
\; \; \; \; \; \; \partial^m_a  \Lambda^\prime_{i\alpha} = 0
\label{A3}
\end{equation}
in coordinate representation.  These derivatives could be written more 
precisely as $(\partial^m_a)_{\Lambda^\prime}$  to show that they act on 
the bosons as usual, but at fixed $\Lambda^\prime$.

The statement (A.2c) is equivalent to the fact that  $S^m_a$ and $T^m_a$ are 
flat connections
\bs
\begin{eqnarray}
\partial^m_a S^n_b - \partial^n_b S^m_a + [S^m_a, S^n_b] & = & 0
\label{A4a} \\
\partial^m_a T^n_b - \partial^n_b T^m_a + [T^m_a, T^n_b] & = & 0 
\label{A4b}
\end{eqnarray}
\es
which follows directly from the properties (2.2) of $\psi_a^i$.  Moreover, 
we find that both connections are divergence-free
\begin{equation}
\partial^m_a S^m_a=  \partial^m_a T^m_a = 0.
\label{A5}
\end{equation}
To see this for $T^m_a$, follow the steps
\bs
\begin{eqnarray}
\partial^m_a (T^m_a)_{ij}& = & \frac{1}{2} \partial^m_a  [\psi_b^i 
\partial^m_a \psi_b^j  - (i \leftrightarrow j)]
\label{A6a} \\
& = & \frac{1}{2} [\psi_b^i \Delta \psi_b^j - (i \leftrightarrow  j)]
\label{A6b} \\
& = & 0
\label{A6c}
\end{eqnarray}
\es
where we used (E.5) in the last step, and similarly for $S^m_a$.  It also 
follows that
\begin{equation}
\partial^m_a F^n_b - \partial^n_b F^m_a + i [F^m_a, F^n_b] = 0, \; \; \; 
\; \; \; \partial^m_a F^m_a= 0
\label{A7}
\end{equation}
so the current $F^m_a$ in eq.~(A.1b) is also a flat divergenceless connection.

The canonical transformation (4.7) or (A.1) can also be understood in terms of 
a unitary (but not gauge-invariant) transformation $K (\phi, 
\Lambda^\prime(\phi))$ 

\bs
\begin{eqnarray}
\pi^{\prime m}_a & = & K^{-1} \pi^m_a K
\label{A8a} \\
\Lambda^\prime_{i\alpha}(\phi) & = &  K^{-1} 
\Lambda^\prime_{i\alpha}(\phi_0) K
\label{A8b} \\
\Lambda^\prime_{i\alpha}(\phi_0)& = & \psi_a^i (\phi_0) \Lambda_{a\alpha}
\label{A8c} \\
\partial^m_a K & = & i K F^m_a, \; \; \; \; \; \; K(\phi_0) = 1
\label{A8d} \\
K & = & P \exp [i \int_{\phi_0}^\phi  d\phi^\prime \cdot F(\phi^\prime, 
\Lambda^\prime(\phi^\prime))]
\label{A8e}
\end{eqnarray}
\es
where $\phi_0$ is a reference point of $\phi$ and $\Lambda_{a\alpha}$ are 
the original constant but not gauge-invariant fermions.  The path-ordered 
operator $K$ is well defined because the current $F^m_a$ is a (divergenceless) 
flat connection.

The gauge-invariant states $\mid \Lambda^\prime\, \rangle$ satisfy
\begin{equation}
\pi^{\prime m}_a \mid \Lambda^\prime(\phi) \, \rangle =  (-i \partial^m_a + 
F^m_a) \mid \Lambda^\prime(\phi) \, \rangle = 0
\label{A9}
\end{equation}
and so may be written in terms of the original fermions as
\begin{equation}
\mid \{\Lambda^\prime_{i\alpha}(\phi)\} \, \rangle =  K^{-1} \mid 
\{\Lambda^\prime_{i\alpha}(\phi_0) = \psi_a^i (\phi_0) \Lambda_{a\alpha}\} 
\, \rangle 
\label{A10}
\end{equation}
although neither factor on the right is separately gauge invariant.

To see the cancellation (4.10) of fermionic terms in the gauge generator 
$G_a$, use eq.~(E.2) for $\partial^m_a \psi_b^i$  to verify the 
intermediate steps
\bs
\begin{eqnarray}
\epsilon_{abc} \phi^m_b \partial^m_c \psi_d^i = \epsilon_{abd} \psi_b^i
\label{A11a} \\
\epsilon_{abc} \phi^m_b (S^m_c)_{df} =  \epsilon_{adf}
\label{A11b}
\end{eqnarray}
\es
where $S^m_c$ is the flat connection in (A.1).  The result in (A.11a) says that 
$\psi_a^i$ transforms in the adjoint of the gauge group.

Similarly, the form (4.12) of the rotation generators follows with the steps
\bs
\begin{eqnarray}
\phi^{[m}_a \partial^{n]}_a \psi_b^j & = & 0
\label{A12a} \\
\phi^{[m}_a (T^{n]}_a)_{ij} & = & \psi_b^i \phi^{[m}_a \partial^{n]}_a 
\psi_b^j= 0
\label{A12b}
\end{eqnarray}
\es
where (A.12a) says that $\psi_a^i$ are singlets under spin$(9)$.

When we make the substitution (A.1) in the Hamiltonian, we encounter
\begin{equation}
\frac{1}{2} \pi^m_a \pi^m_a =\frac{1}{2} \pi^{\prime m}_a \pi^{\prime m}_a 
- F^m_a \pi^{\prime m}_a + \frac{1}{2} F^m_a F^m_a - 
\frac{1}{2}[\pi^{\prime m}_a, F^m_a]
\label{A13}
\end{equation}
where the $F^2$ terms from the shift are quartic in the gauge-invariant 
fermions.  The last term in (A.13) vanishes, however, because the flat 
connection $T^m_a$ has zero divergence.

The explicit form of the shift term in the second canonical transformation 
(4.16) is
\begin{equation}
G^m_a = \frac{1}{2i} (\Lambda^\prime_1 (\Gamma_3 \partial^m_a \Gamma_3) 
\Lambda^\prime_1) = \frac{i}{2} (\Lambda^{\prime\prime}_1 (\Gamma_3 
\partial^m_a \Gamma_3) \Lambda^{\prime\prime}_1)
\label{A14}
\end{equation}
where $\Lambda^{\prime\prime}_i$ are the final gauge-invariant fermions.  
Further details of the derivatives of the matrices $\Gamma_i$ can be found in 
Appendix E.  Here again we find that $\pi^{\prime\prime}$ and $\phi$ are 
canonical variables independent of $\Lambda^{\prime\prime}$.  Moreover, as 
above, the statement $[\pi^{\prime\prime m}_a, \pi^{\prime\prime n}_b] = 0$ 
is equivalent to the fact that $\Gamma_3 \partial^m_a \Gamma_3$ is a flat 
connection, and using eq.~(E.13) we find that this connection is also divergence 
free.

The final form (4.17) of the gauge generators $G_a$ is obtained because 
the shift term $G^m_a$ does not contribute to $G_a$.  To see this, use 
eq.~(E.11) to verify explicitly that $\Gamma_i$ is gauge invariant
\begin{equation}
\epsilon_{abc} \phi^m_b \partial^m_c \Gamma_i = 0
\label{A15}
\end{equation}
and hence that $\epsilon_{abc} \phi^m_b G^m_c = 0$.

We found the following identities helpful
\bs
\begin{eqnarray}
\Gamma_3\phi^{[m}_a \partial^{n]}_a \Gamma_3 & = & - \Gamma_3 \Gamma^{[m} 
\eta^{n]}_3
\label{A16a } \\
\Gamma_3 \Sigma^{mn} \Gamma_3 + i \Gamma_3 \Gamma^{[m} \eta^{n]}_3 & = &  
\Sigma^{mn}
\label{A16b}
\end{eqnarray}
\es
in obtaining the form (4.18) of the rotation generators.  Here $\eta^m_i$ are 
the gauge-invariant angular variables defined in (2.3).

\vskip 1.0cm
\setcounter{equation}{0}
\def\theequation{B.\arabic{equation}}
\boldmath
\noindent{\bf Appendix B: Gauge-Invariant Formulation}
\unboldmath
\vskip 0.5cm

Given the gauge-invariant fermions $\Lambda^\prime$ of Section 4.1 and the 
additional gauge-invariant (but not canonical) coordinates and momenta
\bs
\begin{equation}
\phi^m_i \equiv \psi_a^i \phi^m_a, \; \; \; \; \; \; 
\pi^{\prime m}_i \equiv \psi_a^i \pi^{\prime m}_a = -i \hbar \psi_a^i 
\partial^m_a= -i \hbar D^m_i
\label{B1a}
\end{equation}
\begin{equation}
[\pi^{\prime m}_i, \Lambda^\prime_{j\alpha}] = 0 
\label{B1b}
\end{equation}
\es
$(\pi^{\prime m}_a$ is the independent momentum in (4.7)) we can rewrite the 
supercharges and the Hamiltonian of $SU(2)$ matrix theory entirely in terms of 
gauge-invariant quantities.

Using the derivative formulas of Appendix E, the results are
\bs
\begin{eqnarray}
Q_\alpha & = & (\Gamma^m \Lambda^\prime_i)_\alpha \pi^{\prime m}_i - i \hbar 
\sum_{i \neq j} (\lambda_i \Gamma_i \Lambda^\prime_j)_\alpha 
\frac{(\Lambda^\prime_i \Lambda^\prime_j)}{\lambda_i^2 - \lambda_j^2} 
\nonumber \\
& & \hspace{.25in} + \, \frac{g}{2} \epsilon_{ijk} (\lambda_i \Gamma_i \lambda_j 
\Gamma_j \Lambda^\prime_k)_\alpha 
\label{B2a} \\
H & = & \frac{1}{2} \pi^{\prime m}_i \pi^{\prime m}_i + \frac{g^2}{2} 
(\lambda_1^2 \lambda_2^2 + \lambda_1^2 \lambda_3^2 +\lambda_2^2 
\lambda_3^2) \nonumber \\
& & \hspace{.25in} + \, \frac{i \hbar}{2} \sum_{i \neq j} \frac{1}{\lambda_i^2 - 
\lambda_j^2} \{\phi^m_j \pi^{\prime m}_j + ([\Lambda^\prime_i, 
\Lambda^\prime_j]) \phi^m_i \pi^{\prime m}_j\} \nonumber \\
& & \hspace{.25in} - \, \frac{\hbar^2}{4} \sum_{i 
\neq j} (\Lambda^\prime_i \Lambda^\prime_j)^2 \frac{\lambda_i^2 + 
\lambda_j^2}{(\lambda_i^2 - \lambda_j^2)^2} + \frac{ig \hbar}{2} 
\epsilon_{ijk} (\Lambda^\prime_i \lambda_k \Gamma_k \Lambda^\prime_j)
\label{B2b}
\end{eqnarray}
\es
where $(\Lambda^\prime_i B \Lambda^\prime_j) = \Lambda^\prime_{i\alpha} 
B_{\alpha \beta} \Lambda^\prime_{j \beta}$ and $([\Lambda^\prime_i, 
\Lambda^\prime_j])= \Lambda^\prime_{i\alpha} \Lambda^\prime_{j \alpha} - 
\Lambda^\prime_{j \alpha} \Lambda^\prime_{i\alpha}$.  As expected, $Q$ and 
$H$ are respectively cubic and quartic in the gauge-invariant fermions, 
and the quartic term in the Hamiltonian is just the $F^2$ term of the shift.  
The last term in the Hamiltonian is the Yukawa term $H_F$, whose diagonalization 
is discussed in Appendix C.  

Using chain rules, the gauge-invariant momenta $\pi^{\prime m}_i$ can be 
evaluated explicitly when operating on general gauge-invariant bosonic functions 
$f(\lambda, \eta)$, where the $\eta$ variables are given in (2.3).  The 
results for $\pi^{\prime m}_i$ and the bosonic Laplacian on $f(\lambda, \eta)$, 
\bs
\begin{eqnarray}
- \hbar^2 \Delta & = & \pi^{\prime m}_i \pi^{\prime m}_i + i \hbar 
\sum_{i \neq j} \frac{1}{\lambda_i^2 - \lambda_j^2} \phi^m_j \pi^{\prime m}_j 
\label{B3a} \\
& = & - \hbar^2(\Delta_\lambda + \Delta_\eta) 
\label{B3b}
\end{eqnarray}
\es
are given explicitly in Appendix G.  Here $\Delta_\lambda$, which contains 
the $\lambda$ derivatives, is given in (2.4b) and $\Delta_\eta$ contains the 
$\eta$ derivatives.

We also mention some alternate gauge-invariant forms for the supercharges, 
\bs
\begin{eqnarray}
Q_\alpha & = & (\Gamma^m (\pi^m_a + i \Theta \partial^m_a \sqrt{W}) 
\Lambda_a)_\alpha 
\label{B4a} \\
& = & (\Gamma^m \Lambda^\prime_i)_\alpha \pi^{\prime m}_i - i \hbar 
\sum_{i \neq j} (\lambda_i \Gamma_i \Lambda^\prime_j)_\alpha 
\frac{(\Lambda^\prime_i \Lambda^\prime_j)}{\lambda_i^2 - \lambda_j^2} 
\nonumber \\
& & \hspace{.25in} + \, i (\Gamma^m \Theta \Lambda^\prime_i)_\alpha D^m_i \sqrt{W}.
\label{B4b}
\end{eqnarray}
\es  
Here, $W = g^2 \det\Phi$ is the Claudson-Halpern variable and $\Theta$ is the 
gauge-invariant matrix 
\begin{equation}
\Theta = -i \Gamma_1\Gamma_2 \Gamma_3 
\label{B5}
\end{equation}
which satisfies $\Theta^2 = 1$.  Still another form is
\bs
\begin{eqnarray}
Q_\alpha & = & (\Gamma^m \frac{\phi^m_i}{\lambda_i^2} 
\Lambda^\prime_i)_\beta C_{i\beta\alpha} - i \hbar \sum_{i \neq j} (\lambda_i 
\Gamma_i \Lambda^\prime_j)_\alpha \frac{(\Lambda^\prime_i \Lambda^\prime_j)}
{\lambda_i^2 - \lambda_j^2}
\label{B6a} \\
C_{i\alpha\beta} & = & [-i \hbar \lambda_i \frac{\partial}{\partial\lambda_i} 
+ i \Sigma^{mn} M_i^{mn} - i \Theta \sqrt{W}]_{\alpha\beta}
\label{B6b} \\
M_i^{mn} & = & \phi^{[m}_i \pi^{\prime n]}_i
\label{B6c}
\end{eqnarray}
\es
where the second term in $C_i$, which is of ``spin-orbit'' form, contains all 
the $\eta$ derivatives in the supercharge.  See Appendix G for further 
details.

\vskip 1.0cm
\setcounter{equation}{0}
\def\theequation{C.\arabic{equation}}
\boldmath
\noindent{\bf Appendix C: Diagonalization of the Yukawa Term}
\unboldmath
\vskip 0.5cm

In this Appendix we discuss the exact diagonalization of the 
Yukawa term $H_F$ in the Hamiltonian, keeping $\hbar = g = 1$.

We begin with the expression (4.5) for $H_F$ in terms of the 
gauge-invariant fermions $\Lambda^\prime$,
\bs
\begin{eqnarray}
H_F & = & -\frac{i}{2} \epsilon_{abc} \Lambda_a \Gamma^m \phi^m_b 
\Lambda_c = \frac{1}{2} \Lambda^\prime_{i\alpha} M_{i\alpha,j\beta} 
\Lambda^\prime_{j\beta}
\label{C1a} \\
& & i, j = 1, 2, 3,  \; \; \; \; \; \; \alpha, \beta = 1 \ldots 16
\label{C1b} \\
& & M = -i \left( \begin{array}{ccc}
0 & \lambda_3\Gamma_3 & -\lambda_2\Gamma_2 \\
-\lambda_3\Gamma_3 & 0 & \lambda_1\Gamma_1 \\
\lambda_2\Gamma_2 & -\lambda_1\Gamma_1 & 0
\end{array} \right)
\label{C1c}
\end{eqnarray}
\es
where we have noted that $\epsilon_{abc} \psi^i_b \psi^j_c = 
\epsilon_{ijk} \psi^k_a$ because the eigenvector $\psi$ is a group element 
in the adjoint of $SU(2)$.  The gauge-invariant matrix $M$ is hermitian and 
imaginary, which means that its eigenvalues $\mu$ are real and occur in 
$\pm$ pairs:  if $U$ is one of the 48 eigenvectors of $M$ with eigenvalue 
$\mu$, then $U^\ast$ is also an eigenvector, with eigenvalue $-\mu$.

The matrix $M$ also satisfies
\bs
\begin{equation}
(M^2)_{ij} = (r^2 - 2 \lambda_i^2) \delta_{ij} + \lambda_i\Gamma_i 
\lambda_j\Gamma_j 
\label{C2a}
\end{equation}
\begin{equation}
[M(M^2 - r^2)]_{ij} = 2 \lambda_1 \lambda_2 \lambda_3 \Theta \delta_{ij}
\label{C2b}
\end{equation}
\es
where we have defined $r^2 = \lambda_1^2 + \lambda_2^2 + \lambda_3^2$ and
\begin{equation}
\Theta = -i \Gamma_1 \Gamma_2 \Gamma_3.
\label{C3}
\end{equation}
The gauge-invariant matrix $\Theta$ (which occurs throughout this paper) is 
hermitian and squares to one.  Then (C.2b) gives us a sixth-order algebraic 
equation for the eigenvalues of $M$,
\begin{equation}
[\mu (\mu^2- r^2)]^2 = 4W
\label{C4}
\end{equation}
where $W = (\lambda_1 \lambda_2 \lambda_3)^2 = \det (\Phi)$ is the 
Claudson-Halpern variable.  The solutions of this algebraic equation are 
six real numbers, in three $\pm$ pairs, so that each eigenvalue is 8-fold 
degenerate.  

Furthermore, $\Theta$ commutes with all of the $\Gamma_i$, and hence with the 
matrix $M$, so we can label the eigenvectors of $M$ by their $\Theta$ 
eigenvalues $\pm~1$.  From equation (C.4) above, we find that the three 
eigenvalues $\mu_k, \; k=1,2,3$, corresponding to the $+1$ eigenvalue of
$\Theta$ satisfy
\bs
\begin{equation}
r \leq \mu_3 \leq \frac{2r}{\sqrt{3}}, \; \; \; \; \; \;
-r\leq \mu_2 \leq -\frac{r}{\sqrt{3}}, \; \; \; \; \; \;
- \frac{r}{\sqrt{3}} \leq \mu_1 \leq 0 
\label{C5a}
\end{equation}
\begin{equation}
 \mu_1 + \mu_2 + \mu_3 = 0
\label{C5b}
\end{equation}
\es
and the roots $-\mu_k$ correspond to the $-1$ eigenvalue of $\Theta$.

The origin of the linear relation (C.5b) is as follows: The algebraic 
equation (C.4) gives the eigenvalues as functions of the gauge-invariant 
$\lambda$'s, but in fact only two combinations out of three occur, so that 
$\mu_k = \mu_k(r,W)$.  We also note for use below that the positive 
eigenvalue $\mu_3$ in (C.5a) behaves as
\begin{equation}
\mu_3= R + \frac{(\lambda_1 +\lambda_2)^2}{2R} + \ldots = R + O 
(R^{-2}) 
\label{C6}
\end{equation}
for large $R = \lambda_3$ and $\lambda_1, \lambda_2 = O(R^{-\frac{1}{2}})$.

We are now ready to be more explicit about the eigenfunctions of $M$, 
which may be labelled as
\bs
\begin{eqnarray}
M_{i\alpha, j\beta} U_{j\beta}^{k\nu} = & 
\mu_k U_{i\alpha}^{k\nu}, & \; \; \; \; \; \; 
\Theta_{\alpha\beta} U_{i\beta}^{k\nu} = + U_{i\alpha}^{k\nu} 
\label{C7a} \\
M_{i\alpha, j\beta} U_{i\beta}^{k\nu\ast} = &
-\mu_k U_{i\alpha}^{k\nu\ast}, & \; \; \; \; \; \; 
\Theta_{\alpha\beta} U_{i\beta}^{k\nu\ast}= - U_{i\alpha}^{k\nu\ast}
\label{C7b} 
\end{eqnarray}
\begin{equation}
k=1, 2, 3; \; \; \; \; \; \nu = 1 \ldots 8.
\label{C7c}
\end{equation}
\es
These eigenvectors $U$ and $U^\ast$ form a complete orthonormal set 
\bs
\begin{eqnarray}
U_{i\alpha}^{k\nu\ast} U_{i\alpha}^{k^\prime \nu^\prime} = &
\delta_{kk^\prime} \delta_{\nu\nu^\prime}, & \; \; \; \; \; \; \; \; 
U_{i\alpha}^{k\nu} U_{i\alpha}^{k^\prime \nu^\prime} = 0
\label{C8a} \\
U_{i\alpha}^{k\nu\ast} U_{j\beta}^{k\nu} = &
\delta_{ij} {\displaystyle(\frac{1-\Theta}{2})_{\alpha\beta}}, & \; \; \; \; \; \;
U_{i\alpha}^{k\nu} U_{j\beta}^{k\nu\ast} = \delta_{ij} 
(\frac{1+\Theta}{2})_{\alpha\beta} 
\label{C8b}
\end{eqnarray}
\es
so we can use them to define gauge-invariant creation and annihilation 
operators 
\bs
\begin{equation}
\Lambda^\prime_{i\alpha} = \sum_{k,\nu} (U_{i\alpha}^{k\nu} a^+_{k\nu} + 
U_{i\alpha}^{k\nu\ast} a_{k\nu})
\label{C9a}
\end{equation}
\begin{equation}
\{a_{k\nu}, a^+_{k^\prime \nu^\prime}\} = \delta_{kk^\prime} 
\delta_{\nu\nu^\prime} 
\label{C9b}
\end{equation}
\es		
and we emphasize the pivotal role of the matrix $\Theta$ in the separation into 
creation and annihilation terms.

With this expansion, the original Yukawa term is completely diagonalized
\begin{equation}
H_F = - \sum_{k,\nu} \mu_k a^+_{k\nu} a_{k\nu}	
\label{C10}
\end{equation}
and this is the main result of this Appendix.  Defining $\mid \tilde{0} \, 
\rangle$ by $a_{k\nu} \mid \tilde{0} \, \rangle = 0$ as usual, we find that the state 
with the lowest fermionic energy $H_F \Longrightarrow E_0^F$ is 
\bs
\begin{equation}
(\prod_{\nu=1}^8 a^+_{3\nu}) \mid \tilde{0} \, \rangle \; \; : \; \; E_0^F = -8 
\mu_3
\label{C11a}
\end{equation}
\begin{equation}
E_0^F = -8R + O(R^{-2})	
\label{C11b}
\end{equation}
\es
and we note that the asymptotic form of this energy is the negative of the 
bosonic energy $E_0(R)$ in (2.18).

In this case, one can also make a canonical transformation to independent 
canonical momenta $\tilde{\pi}^m_a$ which commute with the fermion 
creation and annihilation operators,
\bs
\begin{equation}
\tilde{\pi}^m_a = \pi^m_a + \frac{1}{2} \Lambda^\prime_{i\alpha}
(R^m_a)_{i\alpha,j\beta} \Lambda^\prime_{j\beta}
\label{C12a} 
\end{equation}
\begin{equation}
(R^m_a)_{i\alpha,j\beta} = i \sum_{k,\nu} (\partial^m_a 
U^{k\nu\ast}_{i\alpha} U^{k\nu}_{j\beta} + \partial^m_a 
U^{k\nu}_{i\alpha} 
U^{k\nu\ast}_{j\beta})
\label{C12b}
\end{equation}
\begin{equation}
[\tilde{\pi}^m_a, a_{k\nu}] = [\tilde{\pi}^m_a, a^+_{k\nu}] = 0
\label{C12c}
\end{equation}
\es
where $R^m_a$ is again a flat connection.

We have used this transformation and the decomposition
\bs
\begin{equation}
U^{k\nu}_{ia}= u^k_i(\lambda) (\Gamma_i)_{\alpha\beta} \chi^\nu_\beta, \; \; 
\; \; \; \;
U^{k\nu\ast}_{i\alpha} = u^k_i(\lambda) (\Gamma_i)_{\alpha\beta} 
\chi^{\nu\ast}_\beta
\label{C13a}
\end{equation}
\begin{equation}
\sum_i u^k_i(\lambda) u^{k^\prime}_i(\lambda) = \delta^{kk^\prime}
\label{C13b}
\end{equation}
\begin{equation}
\Theta \chi^\nu = + \chi^\nu, \; \; \; \; \; \; \Theta \chi^{\nu\ast} = - 
\chi^{\nu\ast}
\label{C13c}
\end{equation}
\es
to study the rotational properties of the states in this fermionic Hilbert 
space.  The explicit form of the functions $u^k_i(\lambda)$ is easily 
obtained, but is not needed here.  The rotation generators take the form
\bs
\begin{eqnarray}
J^{mn} & = & \tilde{\pi}^{[m}_a \phi^{n]}_a + \frac{i}{2} \sum_{k,\nu,\nu^\prime} 
[a_{k\nu}, a^+_{k\nu^\prime}] \chi^{\nu\ast} (D^{mn} + i \Sigma^{mn}) 
\chi^{\nu^\prime} 
\label{C14a} \\
D^{mn} & = & \phi^{[m}_a \partial^{n]}_a
\label{C14b}
\end{eqnarray}
\es
in this case, and the following list collects the states which are 
singlets \linebreak $(J^{mn} = 0)$ under spin$(9)$:
\bs
\begin{equation}
\mid \tilde{0} \, \rangle e^{-\frac{3\omega}{2}}, \; \; \; \; \; \;
A_k \mid \tilde{0} \, \rangle e^{-\frac{\omega}{2}}
\label{C15a}
\end{equation}
\begin{equation}
A_k A_{k^\prime} \mid \tilde{0} \, \rangle e^{+\frac{\omega}{2}}, \; \; \; \; \; 
\; A_1 A_2 A_3 \mid \tilde{0} \, \rangle e^{+\frac{3\omega}{2}}.
\label{C15b}
\end{equation}
\es
Here we have defined
\begin{equation}
A_k \equiv \prod_{\nu=1}^8 a^+_{k\nu}, \; \; \; \; \; \;
\omega^m_a \equiv \chi^{\nu\ast}_{\alpha} \partial^m_a \chi^\nu_\alpha = 
\partial^m_a \omega
\label{C16}
\end{equation}
and the last relation follows because $\omega^m_a$ is a flat connection.  
The ``lowest'' state (C.11a) appears in this list, and, owing to (C.11b), 
this set of states may provide an alternative description of the spin$(9)$ 
singlet ground state candidate obtained in Section 7.

\vskip 1.0cm
\setcounter{equation}{0}
\def\theequation{D.\arabic{equation}}
\boldmath
\noindent{\bf Appendix D: Assessment of Terms in the Hamiltonian}
\unboldmath
\vskip 0.5cm

Here we examine individual terms, or groups of terms, in the 
transformed Hamiltonian (4.19) and note for each:
\begin{enumerate}
\item its selection rule with respect to the fermion number operator 
\begin{equation}
N_F = \Sigma a_\alpha^+ a_\alpha ~;
\label{D1}
\vspace{-6pt}
\end{equation}
\item its order of magnitude in powers of $R$, using $\lambda_3 = R$ 
and the fact that $\lambda_1$ and $\lambda_2$ are of order 
$R^{-\frac{1}{2}}$ at large $R$;
\vspace{-6pt}
\item its contribution to the asymptotic computation, keeping only 
terms through $O(R^{-2})$ in the effective Hamiltonian.
\end{enumerate}
The details below are given for the ``first computation'' of the text, and 
comments are added at the end which discuss the changes needed for the 
second computation (which allows $\eta_3$ dependence in the reduced state 
vector).

In this discussion, we will use the shorthand $PHP, QHQ, PHQ$ and $QHP$ 
for the various terms in the basic equations, where $PHP$ refers to 
$\langle \, {\bf \cdot} \mid H \mid {\bf \cdot} \, \rangle$ in (5.8), $QHQ$ refers to 
$Q \; H \mid \Psi_Q \, \rangle$ with the ansatz (5.16) for $\mid \Psi_Q \, 
\rangle$, etc.  In this language it will be helpful to state in advance the 
large $R$ systematics
\bs
\begin{eqnarray}
H & = & O(R)
\label{D2a} \\
PHP & = & O(R^{-2})
\label{D2b} \\
QHP, PHQ & = & O(R^{-\frac{1}{2}})
\label{D2c} \\
QHQ & = & O(R)
\label{D2d}
\end{eqnarray}
\es
which we will verify below.  These orders of magnitude (and the fact that 
$P\mid \Psi_Q \, \rangle = 0$) tell us that
\begin{equation}
Q(H - E)\mid \Psi_Q \, \rangle = (H-E) \mid \Psi_Q \, \rangle + 
O(R^{-\frac{1}{2}})
\label{D3}
\end{equation}
and since (as explained in the text) we are only interested in the order 
$R$ 
contributions to these terms, the asymptotic results given here for $QHQ$ 
come entirely from the first term of (D.3).

Finally, it will be useful to note that the shift terms $F^m_a$ in (4.20c) 
and their squares can be written as
\bs
\begin{eqnarray}
F^m_a & = & (a^+ \Gamma_3 a) (T^m_a)_{12} + i(\Lambda_1 \Gamma_3 
\Lambda_3) (T^m_a)_{13} + i(\Lambda_2\Lambda_3) (T^m_a)_{23} \; \; \; \;
\label{D4a} \\
F^m_a F^m_a & = & (a^+ \Gamma_3 a)^2 U_{12} - (\Lambda_1 \Gamma_3 
\Lambda_3)^2 U_{13} - (\Lambda_2 \Lambda_3)^2 U_{23} \; \; \; \;
\label{D4b}
\end{eqnarray}
\es
where $U_{ij}$ is defined in (E.6) and we have used (E.10) to verify that 
there are no cross terms in (D.4b).

\vspace{12pt}
\noindent 1.  $H_B$ and the first term of $H_F$: \\
\begin{equation}
H_0 + H_1 + \lambda_3(N_F -8)
\label{D5}
\end{equation}
where $H_B = H_0 + H_1$ is the bosonic Hamiltonian in (4.19).  The decomposition
of $H_B$ is given in (2.17), now written in terms of independent bosonic 
derivatives.  This group of terms is $O(R)$ and diagonal in $N_F$, but the 
terms of order $R$ cancel in $HP$ because
\begin{equation}
(H_0 -8 R) u_R= 0. ~
\label{D6}
\end{equation}
The $O(R^{-2})$ contributions of these terms to $PHP$ are the derivative 
terms $(\frac{d^2}{dR^2}, \frac{d}{dR})$ in (5.10), as in the bosonic computation 
of Section 2.3.  The contribution of these terms to $QHP$ and $PHQ$ are 
negligible in this computation.

For $QHQ$, it is important to note first that $H_1 = O(R^{-2})$, so these 
terms can be ignored in the present computation.  We find that the 
remaining terms contribute the $O(R)$ terms which are the first four 
terms on the left of each of (5.17a,b), plus the $R$ terms and half of the $U$ 
terms:  The term $\frac{1}{2} U$ comes from the operation of $\Delta$ on each 
$\Gamma_i$ in $\mid \Psi_Q \, \rangle$ (using (E.12)), while the $R$ term 
follows from the $\lambda_3 N_F$ term of (D.5) and the fact that 
$\mid \Psi_Q \, \rangle$ has $N_F = 1$.  The $RD$ terms come from $\Delta$ 
acting as one derivative on the $f$'s and one derivative on $u_R$.  Other 
``cross derivatives'' vanish by virtue of (E.14).

\vspace{12pt}
\noindent 2.  The second and third terms of $H_F$: \\
\begin{equation}
i (\Lambda_2 \Gamma_1 \Lambda_3) \lambda_1 + i (\Lambda_3 \Gamma_2 
\Gamma_3 
\Lambda_1) \lambda_2 
\label{D7}
\end{equation}
This operator changes $N_F$ by $+1$ or $-1$ and is of order 
$R^{-\frac{1}{2}}$.  It gives the entire asymptotic contribution to 
$QHP$ (and to $PHQ$) for our calculation and is written out in equation (5.14).

\vspace{12pt}
\noindent 3.  The first term in $- F^m_a \pi^m_a$: \\
\begin{equation}
i (a^+ \Gamma_3 a) (T^m_a)_{12} \partial^m_a
\label{D8}
\end{equation}
This is diagonal in $N_F$ and of order $R$.  However, it is zero when
acting on $P$, owing to (E.9).  Its only significant contribution is in 
$QHQ$, where it acts upon the matrices $\Gamma_1$ and $\Gamma_2$, according to 
(E.15).  This term exchanges the fermion bilinears
\begin{eqnarray}
\lefteqn{(T^m_a)_{12} \partial^m_a [(a^+ \Gamma_3 a) 
\left\{ \begin{array}{c}
(a^ + \Gamma_1 \Lambda_3) \\ (\Lambda_3 \Gamma_2 \Gamma_3 a^+)
\end{array} \right\} \mid 0 \, \rangle ]} \hspace{2in} \nonumber \\
& & \hspace{-.5in} = Z \left\{ \begin{array}{c}
(\Lambda_3 \Gamma_2 \Gamma_3 a^+) \\ -(a^ + \Gamma_1 \Lambda_3)
\end{array} \right\} \mid 0 \, \rangle
\label{D9}
\end{eqnarray}
to leading order in $R$ and produces the ``mixing'' terms in eqs.~(5.17) 
proportional to $Z$.

\vspace{12pt}
\noindent 4.  The second and third terms in $- F^m_a \pi^m_a$: \\
These terms (see (D.4a)) raise or lower $N_F$ by one and are zero acting on 
$P$ (see (E.9)); they are too small to make any contribution to the present 
calculation.

\vspace{12pt}
\noindent 5.  The term $-G^m_a \pi^m_a$: \\
This gives zero in $PHP$ by (E.14) and is too small to contribute 
elsewhere.

\vspace{12pt}
\noindent 6.  The first term in $\frac{1}{2} F^m_a F^m_a$: \\
This term (see (D.4b)), is diagonal in $N_F$ but zero when acting on $P$.  A 
useful fact here is
\begin{equation}
(a^+ \Gamma_3 a)^2 a^+_\alpha \mid 0 \, \rangle= a^+_\alpha \mid 0 \, \rangle
\label{D10}
\end{equation}
and the asymptotic contribution $\frac{1}{2} U_{12} \mid \Psi_Q 
\, \rangle$ is obtained for this term in $Q H \mid \Psi_Q \, \rangle$.  
This gives the remaining half of the $U$ terms in (5.17).

\vspace{12pt}
\noindent 7.  The second and third terms in $\frac{1}{2} F^m_a F^m_a$: \\
These contribute to $PHP$ as
\begin{equation}
\langle \, 0 \mid \frac{1}{2} F^m_a F^m_a \mid \; 0 \; \, \rangle = 2(U_{13} + 
U_{23}) 
= \frac{4}{R^2} + \ldots
\label{D11}
\end{equation}
and hence make a contribution of $\frac{4}{R^2}$ to (5.9).

\vspace{12pt}
\noindent 8.  The term $\frac{1}{2} G^m_a G^m_a$: \\
This contributes to $PHP$ as follows.  Using (E.16) and (E.12) we compute
\begin{eqnarray}
\langle \, 0 \mid \frac{1}{2} G^m_a G^m_a \mid \; 0 \; \, \rangle & = & 
\frac{1}{16} {\rm Trace} [(\partial^m_a \Gamma_3) (\partial^m_a 
\Gamma_3)] = -\frac{1}{16} {\rm Trace}[(\Gamma_3 \Delta \Gamma_3)] 
\nonumber \\
& & = \frac{6}{\lambda_3^2} + U_{13} + U_{23} = \frac{8}{R^2} + \ldots
\label{D12}
\end{eqnarray}
and hence this group of terms contributes $+\frac{8}{R^2}$ to equation (5.9).

\vspace{12pt}
\noindent 9.  The term $F^m_a G^m_a$. \\
Using (E.15), this term is negligible in this calculation.

\vspace{12pt}
\noindent For the second computation, we must allow for the fact that the 
reduced state vector $\mid \psi (R, \Lambda_3, \eta_3) \, \rangle$ is a 
function also of the gauge-invariant angular variable $\eta_3$.  This means 
that we must reexamine those terms above which involve derivatives with respect 
to $\eta_3$, namely $\Delta$ and the shift terms $F\pi$ and $G\pi$.  The 
result for $\Delta$ is discussed in Section 7, and, because derivatives of 
$\eta_3$ are at least one power of $R^{-1}$ smaller than the terms we have kept, 
we find no new contributions from the shift terms.

\vskip 1.0cm
\setcounter{equation}{0}
\def\theequation{E.\arabic{equation}}
\boldmath
\noindent{\bf Appendix E: Derivatives}
\unboldmath
\vskip 0.5cm

We list here a number of useful formulas for the differentiation of 
the bosonic variables introduced in the text.  The notation is
\[\partial^m_a = \frac{\partial}{\partial\phi^m_a}, \; \; \; \; \; \; 
\Delta = \partial^m_a \partial^m_a \]
and we adopt here the generalization
\[m,n = 1 \ldots d; \; \; \; \; \; \; a,b,c = 1 \ldots g; \; \; \; \; \; 
i,j,k = 1 \ldots g \; \; \; \; \; \; (g \leq d) \]
although only $d=9$ and $g=3$ apply for $SU(2)$ matrix theory.

Using the familiar method of matrix-perturbation theory, one derives 
the following two basic formulas for differentiation of $\lambda_i$ and 
$\psi_a^i$, defined in (2.2),
\begin{eqnarray}
\partial^m_a \lambda_i & = & \psi_a^i \psi_b^i \frac{\phi^m_b}{\lambda_i}
\label{E1} \\
\partial^m_a \psi_b^i & = & \sum_{j \neq i} \psi_b^j \phi^m_c 
\frac{\psi_a^i 
\psi_c^j + \psi_c^i \psi_a^j}{\lambda_i^2 - \lambda_j^2}.  	
\label{E2}
\end{eqnarray}
All that follows is derived by repeated application of these relations and 
the prior definitions.

When $f(\lambda)$ is any function of the $\lambda_i$, we have
\begin{eqnarray}
\Delta f(\lambda) & = & \sum_i\{\frac{\partial^2}{\partial\lambda_i^2} + 
[\frac{(d-g)}{\lambda_i} + \sum_{j \neq i} \frac{2\lambda_i} {(\lambda_i^2 
- \lambda_j^2)}] \frac{\partial}{\partial\lambda_i}\}f(\lambda)  
\;\;\;\;\;\;\;\;\;\;\;
\label{E3} \\
\nonumber \\
(\partial^m_a \psi_b^i) (\partial^m_a f(\lambda)) & = & 0
\label{E4} \\
\nonumber \\
\Delta \psi_a^i & = & \psi_a^i \{- \sum_{j \neq i} U_{ij}\}
\label{E5} \\
\nonumber \\
U_{ij} & \equiv & \frac{(\lambda_i^2 + \lambda_j^2)}{(\lambda_i^2 - \lambda_j^2)^2}.
\label{E6}
\end{eqnarray}
The flat matrix connections $T$ were introduced in (4.7) and Appendix A:
\begin{eqnarray}
(T^m_a)_{ij} & = & (\psi_b^i \partial^m_a \psi_b^j) = (1 - \delta_{ij}) 
\phi^m_b \frac{(\psi_b^i \psi_a^j + \psi_a^i \psi_b^j)}{(\lambda_j^2 - 
\lambda_i^2)}
\label{E7} \\
\partial^m_a (T^m_a)_{ij} & = & 0
\label{E8} \\
\nonumber \\
(T^m_a)_{ij}(\partial^m_a f(\lambda)) & = & 0
\label{E9} \\
\nonumber \\
(T^m_a)_{ij}(T^m_a)_{kl} & = & (1 - \delta_{ij})(\delta_{ik} \delta_{jl} - 
\delta_{il} \delta_{jk}) U_{ij}.
\label{E10}
\end{eqnarray}
The gauge-invariant matrices $\Gamma_i$ are defined in (4.6).  They are 
real, symmetric, traceless, anti-commuting and satisfy $(\Gamma_i)^2 = 1$:
\begin{equation}
\partial^m_a \Gamma_i = \frac{\Gamma^n}{\lambda_i} \{\delta_{nm} 
\psi_a^i - \phi^n_b \phi^m_c \psi_b^i \psi_c^i \frac{\psi_a^i}{\lambda_i^2} + 
\phi^n_b \phi^m_c \sum_{j \neq i} \psi_b^j \frac{(\psi_c^i \psi_a^j + \psi_a^i 
\psi_c^j)} {(\lambda_i^2 - \lambda_j^2)}\}
\label{E11}
\vspace{-12pt}
\end{equation}
\begin{eqnarray}
\Delta \Gamma_i & \! \! = \! \! & \Gamma_i \{-\frac{(d-g)}{\lambda_i^2} - 
\sum_{j \neq i} U_{ij}\}	
\label{E12} \\
\nonumber \\
\partial^m_a (\Gamma_i \partial^m_a \Gamma_i) & \! \! = \! \! & 0 
\; \; \; \; \; \; \mbox{(no sum on $i$)}
\label{E13} \\
\nonumber \\
(\partial^m_a \Gamma_i) (\partial^m_a f(\lambda)) & \! \! = \! \! & 0
\label{E14}	\\
\nonumber \\
(T^m_a)_{ij} (\partial^m_a \Gamma_k) & \! \! = \! \! & (1 - \delta_{ij}) 
\frac{2\lambda_k} {(\lambda_i^2 - \lambda_j^2)^2} \{\delta_{kj} \lambda_i 
\Gamma_i - \delta_{ki} \lambda_j \Gamma_j\} 
\label{E15} \\
\nonumber \\
(\partial^m_a \Gamma_i)_{\alpha\beta} (\partial^m_a \Gamma_j)_{\gamma\delta} 
& \! \! = \! \! & \delta_{ij} \{ \frac{[(\Gamma^m)_{\alpha\beta} 
(\Gamma^m)_{\gamma\delta} - \sum_k (\Gamma_k)_{\alpha\beta} 
(\Gamma_k)_{\gamma\delta}]} {\lambda_i^2} \hspace{1in} \nonumber \\
& & \! \! + \! \sum_{k \neq i} U_{ik} (\Gamma_k)_{\alpha\beta} (\Gamma_k)_{\gamma\delta} 
\} \! - \! (1 \! - \delta_{ij}) U_{ij} (\Gamma_j)_{\alpha\beta} (\Gamma_i)_{\gamma\delta}
\; \; \; \label{E16} \\
\Delta (\Gamma_i \Gamma_j) & \! \! = \! \! & (1 - \delta_{ij}) \Gamma_i\Gamma_j 
\{-\frac{(d-g)}{\lambda_i^2} - \sum_{k \neq i} U_{ik} - \frac{(d-g)}{\lambda_j^2}
\nonumber \\
& & \! \! - \! \sum_{k \neq j} U_{jk} + 2U_{ij} \}.  
\label{E17}
\end{eqnarray}

In the case of $SU(2)$, the special gauge-invariant matrix 
\begin{equation}
\Theta = -i \Gamma_1 \Gamma_2 \Gamma_3 = -\frac{i}{6} \epsilon_{abc} \Gamma^m 
\Gamma^n \Gamma^p \frac{\phi^m_a \phi^n_b \phi^p_c}{\lambda_1\lambda_2 
\lambda_3}
\label{E18}
\end{equation}
is imaginary, antisymmetric, traceless, has square equal to the unit 
matrix, and commutes with the matrices $\Gamma_i$:
\begin{eqnarray}
\phi^m_a \partial^m_b \Theta & = & 0
\label{E19} \\
\Delta \Theta & = & \Theta \{-(d-3) \sum_i \frac{1}{\lambda_i^2}\}
\label{E20} \\
\partial^m_a (\Theta \partial^m_a \Theta) & = & 0.
\label{E21}
\end{eqnarray}

\vskip 1.0cm
\setcounter{equation}{0}
\def\theequation{F.\arabic{equation}}
\boldmath
\noindent{\bf Appendix F: Integrals}
\unboldmath
\vskip 0.5cm

When we average over the fast variables $\lambda_1$ and $\lambda_2$ with 
the Gaussian function (2.18a), the following class of two-dimensional integrals 
occur:
\begin{equation}
\int_0^\infty ds \int_0^s dt (s^2-t^2) (st)^M (s^2 + t^2)^N \exp(-s^2 - 
t^2)
= \frac{(N + M + 1)!}{(M+1) 2^{M+2}}.
\label{F1}
\end{equation}
This formula gives us useful averages for our asymptotic calculation.  
Using the notation
\begin{equation}
\langle \, f(\lambda_1, \lambda_2) \, \rangle = \int d^2 \lambda \sigma_\infty 
\mid u_R \mid^2 \frac{f(\lambda_1, \lambda_2)}{\int d^2 \lambda \sigma_\infty 
\mid u_R \mid^2}
\label{F2}
\end{equation}
(see eq.~(2.21) and set $\hbar = g = 1$) we find, for general values of $d$,
\begin{eqnarray}
\langle \,\lambda_1^2 +\lambda_2^2\, \rangle & = & \frac{(d-1)}{R}
\label{F3} \\
\langle \,(\lambda_1^2 +\lambda_2^2)^2\, \rangle & = & \frac{d(d-1)}{R^2}
\label{F4} \\
\langle \,\lambda_1 \lambda_2\, \rangle & = & \frac{(d-2)}{2R}
\label{F5} \\
\langle \,\lambda_1^2 \lambda_2^2\, \rangle & = & \frac{(d-1)(d-2)}{4R^2}
\label{F6}
\end{eqnarray}
where $d=9$ for matrix theory.

\vskip 1.0cm
\setcounter{equation}{0}
\def\theequation{G.\arabic{equation}}
\boldmath
\noindent{\bf Appendix G: Gauge-Invariant Angular Variables}
\unboldmath
\vskip 0.5cm

Here we will express the bosonic Laplacian in terms of the complete 
set $(\lambda_i, \eta^m_i)$ of gauge-invariant variables, regarding the 
$\lambda$'s and $\eta$'s respectively as radial and angular variables.  
The result can be arranged in several ways and the one shown below has 
particular advantages for our work.  

We start, as in Appendix B, with the gauge-invariant bosonic variables
\begin{equation}
\phi^m_i \equiv \psi_a^i \phi^m_a
\label{G1}
\end{equation}
and the Lie derivatives (which do not act on the gauge-invariant fermions)
\begin{equation}
i \pi^{\prime m}_i = D^m_i \equiv \psi_a^i \partial^m_a = \psi_a^i 
\frac{\partial}{\partial\phi^m_a} 
\label{G2}
\end{equation}
so that the Laplacian can be written as
\begin{equation}
\Delta = D^m_i D^m_i + (\partial^m_a \psi_a^j) D^m_j = D^m_i D^m_i - 
\sum_{i \neq j} y_{ij} \phi^m_j D^m_j.
\label{G3}
\end{equation}
Here we have made use of (E.2) and
\begin{equation}
y_{ij} \equiv \frac{1}{(\lambda_i^2 - \lambda_j^2)}.
\label{G4}
\end{equation}
The formula
\begin{equation}
D^m_i \phi^n_j= \delta_{ij} [\delta_{mn} + \sum_{k \neq i} y_{ik} \phi^m_k 
\phi^n_k] 
- (1 - \delta_{ij}) y_{ij} \phi^m_j \phi^n_i
\label{G5}
\end{equation}
also follows from results in Appendix E and will be used below.

Next, we write the orbital angular momentum operator $M^{mn}$ in terms of 
these new derivatives,
\bs
\begin{eqnarray}
M^{mn} & = & -i [\phi^m_a \partial^n_a - \phi^n_a \partial^m_a] = \sum_i 
M_i^{mn}
\label{G6a} \\
M_i^{mn} & \equiv & -i [\phi^m_i D^n_i - \phi^n_i D^m_i]
\label{G6b}
\end{eqnarray}
\es
and note that the operators $M_i^{mn}$ are hermitian in the measure 
$(d\phi)$, 
although their algebra is not simple.  Now calculate the trace of the 
square of each $M_i$:
\bs
\begin{eqnarray}
M_i^2 & = & - \sum_{m<n} [\phi^m_i D^n_i - \phi^n_i D^m_i]^2
\label{G7a} \\
& = & - \lambda_i^2 (D^n_i)^2 + (\phi^m_i D^m_i)^2 + [d-2 + \sum_{k \neq i} 
y_{ik} \lambda_k^2] \phi^m_i D^m_i.
\label{G7b}
\end{eqnarray}
\es
Combining this result with equation (G.3) for the Laplacian, we find
\begin{equation}
\Delta = \frac{1}{\lambda_i^2} \{(\phi^m_i D^m_i)^2 - M_i^2 + [d-2 + 
\sum_{j \neq i} y_{ij}(\lambda_i^2 + \lambda_j^2)] \phi^m_i D^m_i \}
\label{G8}
\end{equation}
and then noting that
\begin{equation}
\phi^m_i D^m_i = \lambda_i \frac{\partial}{\partial\lambda_i} \; \; \; \; 
\; \; \mbox{(no sum on $i$)}
\label{G9}
\end{equation}
we can simplify this formula to the nice form
\bs
\begin{eqnarray}
\Delta & = & \Delta_\lambda + \Delta_\eta
\label{G10a} \\
\Delta_\eta & = & - \sum_i \frac{M_i^2}{\lambda_i^2}
\label{G10b}
\end{eqnarray}
\es
Here $\Delta_\lambda$, which contains the $\lambda$ derivatives, was given 
earlier in eq.~(E.3) and we will see that $\Delta_\eta$, which is negative 
semidefinite, contains only derivatives with respect to the angular 
variables $\eta$
\begin{equation}
\eta^m_i \equiv \frac{\phi^m_i }{\lambda_i}
\label{G11}
\end{equation}
which complement the radial variables $\lambda$.

For any function $f(\lambda, \eta)$, the chain rule gives
\bs
\begin{eqnarray}
D^m_i f & = & \eta^m_i \frac{\partial}{\partial\lambda_i} f + (D^m_i \eta^n_j) 
\partial^n_j f
\label{G12a} \\
D^m_i \eta^n_j & = & \frac{\delta_{ij}}{\lambda_i} [\delta_{mn} - \sum_k 
\eta^m_k \eta^n_k + \sum_{k \neq i} y_{ik} \lambda_i^2 \eta^m_k \eta^n_k] 
\nonumber \\
& & - (1-\delta_{ij}) y_{ij} \lambda_i \eta^m_j \eta^n_i
\label{G12b}
\end{eqnarray}
\es
where we have defined the $\eta$ derivative
\begin{equation}
\partial^m_i \equiv \frac{\partial}{\partial\eta^m_i}, \; \; \; \; \; \;
\partial^m_i \eta^n_j = \delta_{mn} \delta_{ij}
\label{G13}
\end{equation}
and (G.12b) is closely related to (E.11).

Using (G.12), we re-express the operators $M_i^{mn}$ in terms of the 
variables $\lambda$ and $\eta$.  As expected, all $\frac{\partial}
{\partial\lambda_i}$ terms cancel out and we find that

\bs
\begin{equation}
M_i^{mn} = L_i^{mn} + i\sum_{j \neq i} \eta^{[m}_i \eta^{n]}_j \{x_{ij} 
(\eta_i \partial_j) + x_{ji} (\eta_j \partial_i)\}
\label{G14a}
\end{equation}
\begin{equation}
L_i^{mn} \equiv -i \eta^{[m}_i \partial^{n]}_i
\label{G14b}
\end{equation}
\es
where $L_i^{mn}$ is the naive angular momentum operator for the $\eta$ 
variables and
\begin{equation}
x_{ij} \equiv \frac{\lambda_i^2}{(\lambda_i^2-\lambda_j^2)}, \; \; \; \; \; 
\;
(\eta_i \partial_j) \equiv \eta^m_i \partial^m_j.
\label{G15}
\end{equation}
Using the naive $\eta$ derivative in (G.13), it is not difficult to check 
that the operators $M_i$ in (G.14) respect the $\eta$ constraints
\begin{equation}
M_i^{mn} (\eta^p_j \eta^p_k) = 0
\label{G16}
\end{equation}
and it follows directly that the operators $M_i^{mn}$ are hermitian in the 
gauge-invariant measure $d^3\lambda (d^3\eta)$ (see (7.4)).  Taken with (G.14), 
the form of the Laplacian in (G.10) is the central result of this appendix.

For the discussion below, we will also need the form of the operators 
$M_i$ 
\bs
\begin{eqnarray}
M_3^{mn} & = & L_3^{mn} + i\sum_{j = 1,2} \eta^{[m}_3 \eta^{n]}_j 
(\eta_3 \partial_j) + \ldots
\label{G17a} \\
M_1^{mn} & = & -i \{ \eta^{[m}_1 \partial^{n]}_1 - \eta^{[m}_1 \eta^{n]}_2 
[x_{21} (\eta_2 \partial_1) + x_{12} (\eta_1 \partial_2)] \nonumber \\
& &	+ \eta^{[m}_1 \eta^{n]}_3 (\eta_3 \partial_1) \} + i 
(\frac{\lambda_1^2}{R^2}) \eta^{[m}_1 \eta^{n]}_3 (\eta_{[3} \partial_{1]}) 
+ \ldots
\label{G17b} \\
M_2^{mn} & = & -i \{\eta^{[m}_2 \partial^{n]}_2 - \eta^{[m}_2 \eta^{n]}_1 
[x_{12} 
(\eta_1 \partial_2) + x_{21} (\eta_2 \partial_1)] \nonumber \\
& & + \eta^{[m}_2 \eta^{n]}_3 (\eta_3 \partial_2) \} + i 
(\frac{\lambda_2^2}{R^2}) 
\eta^{[m}_2 \eta^{n]}_3 (\eta_{[3} \partial_{2]}) + \ldots 
\label{G17c}
\end{eqnarray}
\es
in the asymptotic region, $R = \lambda_3 > > \lambda_1, \lambda_2 = 
O(R^{-\frac{1}{2}})$.  The extra term (7.14) of the second computation in 
the text
\bs
\begin{eqnarray}
M_3^{mn} f(\eta_3) & = & L_3^{mn} f(\eta_3) + O(R^{-1})
\label{G18a} \\
-\frac{1}{2} \Delta f(\eta_3) & = & (\frac{L_3^2}{2R^2} + O(R^{-3})) f(\eta_3)
\label{G18b}
\end{eqnarray}
\es
follows immediately from the asymptotic form of $M_3$ in (G.17), the 
$M_{1,2}$ terms failing to contribute at this order.

In what follows, we will use the results above to outline a strategy 
for proving the following conjecture: 

\begin{itemize}
\item[a)] the eigenvalues $\epsilon$ of the bosonic operator $H_0$ in (2.17)
satisfy 
\begin{equation}
\epsilon \geq (d-1)R = 8R 
\label{G19} 
\end{equation}
and $u_R$ in (2.18) is the only state which 
realizes the minimum.
\vspace{-10pt}
\item[b)] the eigenvalues of the bosonic operator 
\begin{equation}
H_0^\prime = H_0 + \frac{M_1^2}{2 \lambda_1^2} + \frac{M_2^2}{2 
\lambda_2^2}
\label{G20}
\end{equation}
also satisfy $\epsilon \geq (d-1)R = 8R$ and $u_R$ is the only state which 
realizes the minimum.  Here $M_{1,2}$ are given by their leading (first 
four) terms in (G.17b,c).
\end{itemize}
The operator $H_0$ is the dominant part (that is, it contains all terms of 
order $R$) of the bosonic Hamiltonian in the gauge- and rotation-invariant 
sector; it contains only the fast derivatives $\frac{\partial}{\partial
\lambda_{1,2}}$ with the slow variable $R = \lambda_3$ as a parameter.  
The operator $H_0^\prime$ is the dominant part (in the same sense) of the full 
bosonic Hamiltonian $H_B$ at large $R$, including the gauge-invariant angular
excitations; it involves only the fast derivatives $\frac{\partial}{\partial
\lambda_{1,2}}$ and $\frac{\partial}{\partial\eta_{1,2}}$, with the slow 
variables $R = \lambda_3$ and $\eta_3$ as parameters.

If true, this conjecture implies that our $\eta_1, \eta_2$-independent 
projector state \linebreak $\mid {\bf \cdot} \, \rangle$ in (7.3) is the only state whose 
associated effective Hamiltonian (including $-8R$ from the fermions) has no 
linear term in $R$.  

There is strong evidence for (a), though we have not tried to prove it:
It is straightforward to find a large class of radial eigenfunctions 
$u_{m,n} (\lambda_1, \lambda_2)$ of $H_0$ (or $H_0^\prime$) with
\begin{equation}
\epsilon = R(d-1 + 2(m+n)), \; \; \; \; \; \; m,n = 0,1,2 \ldots
\label{G21}
\end{equation}
where $u_{0,0} = u_R$.  Assuming (a), we can prove (b) as follows.  The 
positive semi-definite operators $M_i^2/\lambda_i^2, \; \; \; i = 
1,2$ can only give additional positive semi-definite contributions to 
$\epsilon$, beyond $(d-1)R$.  So to prove (b), we only need to show 
that there are no non-constant solutions to the differential equations 
\begin{equation}
M_i^{mn} {\rm v}(\eta_1, \eta_2) = 0; \; \; \; \; \; \; i = 1,2 \; \; \; \; 
\; \forall mn
\label{G22}
\end{equation}
where the $M$'s are given by their leading terms at large $R$.  We have 
explicitly checked that this is true.


\end{document}